\documentclass[aps, prb, reprint, twocolumn, superscriptaddress, footinbib]{revtex4-2}

\usepackage{geometry}
\usepackage[utf8]{inputenc}

\usepackage{amsmath}
\usepackage{bbm}
\usepackage{bm}
\usepackage{graphicx}
\usepackage[dvipsnames]{xcolor} % Color in text
\usepackage{subfig}
\usepackage[export]{adjustbox}

\usepackage{hyperref}
\usepackage[capitalise]{cleveref} % Loaded amsmath->hyperref->cleveref

\usepackage{comment}

\def\mytitle{Nonlocal Transport of Heat in Equilibrium Drift-Diffusion Systems} % title

\begin{document}

%%%%%%%%%%%%%%%%%Affilitations%%%%%%%%%%%%%%%%%
	
	\author{Florian St{\"a}bler}
		\affiliation{D\'epartement de Physique Th\'eorique, Universit\'e de Gen\`eve, CH-1211 Gen\`eve 4, Switzerland}
	\author{Eugene Sukhorukov}
		\affiliation{D\'epartement de Physique Th\'eorique, Universit\'e de Gen\`eve, CH-1211 Gen\`eve 4, Switzerland}
	
%%%%%%%%%%%%%%%%%Title%%%%%%%%%%%%%%%%%
	
	\title{\mytitle}
	\date{\today}

\begin{abstract}

The amount of heat an integer quantum Hall edge state can carry in equilibrium is quantized in universal units of the heat flux quantum $J_q= \frac{\pi k_B^2}{12 \hbar}T^2$ per edge state.  We adress the question of how heat transport in realistic one dimensional devices can differ from the usual chiral Luttinger liquid theory. We show that a local measurement can reveal a nonquantized amount of heat carried by the edge states, despite a globally equilibrium situation. More specifically, we report a heat enhancement effect in edge states interacting with ohmic reservoirs in the presence of nonlocal interactions or chirality breaking diffusive currents. In contrast to a nonequilibrium, nonlinear drag effect, we report an equilibrium, linear phenomenon. The chirality of the edge states creates additional correlations between the reservoirs, reflected in a higher than quantum heat flux in the chiral channel. We show that for different types of coupling the enhancement can be understood as static or dynamical backaction of the reservoirs on the chiral channel. We show that our results qualitatively hold by replacing the dissipative ohmic reservoirs by an energy conserving mesoscopic capacitor and consider the respective transmission lines for different types of interaction.

\end{abstract}

\maketitle

\section{Introduction}

One dimensional systems, especially quantum Hall (QH) systems provide an interesting platform to build quantum thermoelectric devices due to the chiral nature of its edge states. The theoretical framework of QH edge states contains a lot of seeming paradoxes due to the fact that, on one hand, edge states show many signature properties of one dimensional (chiral) states, like a quantized charge or thermal  Hall conductance \cite{kane_quantized_1997,banerjee_observation_2018} and a quantized heat flux in thermal equilibrium \cite{granger_observation_2009,le_sueur_energy_2010,venkatachalam_local_2012}, but on the other hand it is not intuitively clear, how strong interactions in mesoscopic devices affect this common knowledge.

It was demonstrated, that interactions introduced by coupling the edge states to metallic granulas, show a violation of the perfect quantization \cite{sivre_electronic_2019,duprez_dynamical_2021} of heat if the system is driven out of equilibrium. 
Recently, experiments found a deviation from this perfect quantization due to unknown mechanisms \cite{granger_observation_2009,le_sueur_energy_2010,venkatachalam_local_2012}. It was an open question if dissipation due to the naturally present disorder in the edges can be an explanation for this loss of heat in the edge \cite{goremykina_heat_2019}. This motivated us, in the previous paper, to develop an effective model of a chiral edge in the presence of intrinsic dissipation \cite{stabler_transmission_2022}. We surprisingly found that even in the presence of dissipation of arbitrary strength in equilibrium the energy flux carried by the edge retains the universal value. The dissipated heat is exactly cancelled by the back action effect of the dissipative degrees of freedom on the chiral edge state, guaranteeing perfect quantization.

\begin{figure}[htbp]%
    \centering
    \subfloat[\centering Microscopic image of a disordered edge.]{\includegraphics[width=.95\linewidth,valign=c]{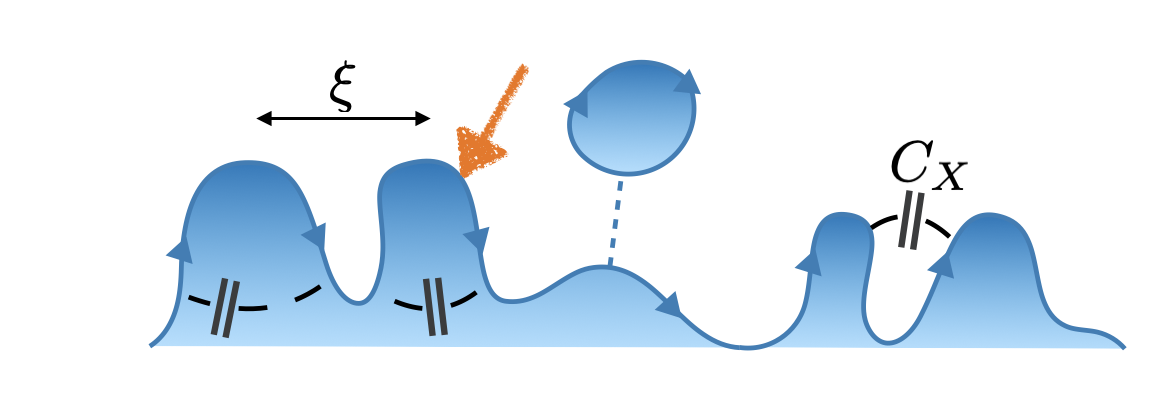}%
    \label{fig:edgepic1}}%
    
    \subfloat[\centering Experimental setup to measure the effective temperature of the edge.]{{\includegraphics[width=.9\linewidth,valign=c]{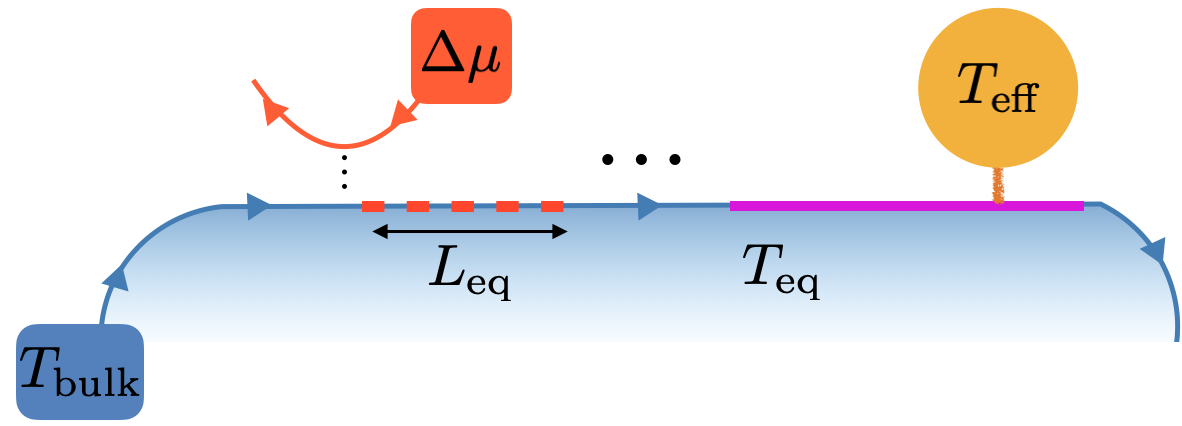} }\label{fig:edgepic2}}%
   \caption{ \cref{fig:edgepic1}: The edge of a quantum hall states can be disordered and form lumps and puddle defects. The disorder is assumed to have a characteristic length scale $\xi$. We explicitly model the interaction between these edge structures and show that a local measurement of the effective temperature (indicated by the orange arrow) could reveal an unexpectedly high temperature. \cref{fig:edgepic2}: The experiment is done by creating a non-equilibrium distribution function in the edge, by mixing two edge states at different chemical potentials at a quantum point contact. After the equilibration length $L_{\text{eq}}$ we assume that the edge retains a local equilibrium temperature $T_{\text{eq}}>T_{\text{bulk}}$. The subject of this paper concerns the effective temperature after the equilibration has taken place and in a part of the edge subject to interactions. As we show $T_{\text{eff}}$ is not necessarily equal to $T_{\text{eq}}$, depending on where and how one measures. Since we are in a local equilibrium, we assume that our system is subject to a boundary current with a correlation function in local equilibrium with the corresponding temperature $T_{\text{eq}}$ for the two node setup or periodic boundary conditions for the transmission line setup. The latter is justified since the collective mode decays and we assume an infinite system.  }
    \label{fig:edgepic}%
\end{figure}

One effect which has been studied in conventional mesoscopic devices is the mutual electron drag effect \cite{narozhny_coulomb_2016,levchenko_coulomb_2008}. A device with an active part, driven out of equilibrium, induces a current in a passive part of the system. Various mechanisms for the drag of charge currents are known, such as combined electron-phonon, electron-photon, and electron-ion interactions \cite{raichev_phonon_2020,strait_revisiting_2019, gurevich_drag_2015}. In contrast to this is the thermal drag effect, where a temperature imbalance in the system induces a heat current, mediated by a Coulomb interaction between the active and passive part of the system. Recently a similar drag effect has been found in a chiral system consisting of two QH edge states interacting cross capacitively through two metallic granulas \cite{idrisov_thermal_2022}. The effect leads to a heat drag if the active part of the system is driven out of equilibrium, but is a linear phenomenon; it does not rely on the rectification of noise due to a nonlinear element, as seen in other similar drag effects.

The idea of crosscapacitive coupling as a form of nonlocal coupling was also realized in a nonchiral system and led to a brownian gyration effect, where the coupling leads to a persistent current, similar to a heat pump or ratchet, where heat was pumped from the hot to the cold reservoir by the current \cite{filliger_brownian_2007,chiang_electrical_2017}.

{\textit{In this paper}} we  focus on QH edges at integer filling factors subject to strong interactions. The goal of this paper is to analyze the reason for the perfect quantization of heat flux in chiral edge states with intrinsic dissipation and show how it can be broken in equilibrium drift-diffusion systems. Our main result is that a local measurement of heat can reveal a nontrivial amount of heat current carried by the edge states in a globally equilibrium system. To understand this statement we start from a physical model of the QH edge.

In real systems, the edge of a QH system is typically disordered by lumps and defects, as shown in  \cref{fig:edgepic1}. We assume that the disorder is characterized by a typical disorder length scale $\xi$ and leads to a decay of a collective mode present in the edge \cite{goremykina_heat_2019,stabler_transmission_2022} with a decay rate proportional to $\xi^{-1}$. We refer to a certain class of experiments with an experimental setup shown in  \cref{fig:edgepic2} \cite{le_sueur_energy_2010,venkatachalam_local_2012,granger_observation_2009}. A quantum point contact (QPC) creates a non-equilibrium distribution function in the edge. We do not study the equilibration in this paper but assume that after some equilibration length $L_{\text{eq}}$ the edge reaches a local thermal equilibrium temperature $T_{\text{eq}}$. We can view this as a boundary condition for the edge state and assume that somewhere upstream of the detector we have a boundary current with a thermal noise power corresponding to the  temperature $T_{\text{eq}}$. Furtermore, the experiments show that the edge is thermally isolated from the bulk, but this does not exclude the possibility that the energy of the collective mode can be dissipated to unknown degrees of freedom, e.g. to the creation of electron-hole excitations in the edge. We will refer to this as intrinsic dissipation.

A first attempt to study this system with a hydrodynamic model of the edge with intrinsic dissipation \cite{goremykina_heat_2019} led to the non-trivial problem that the definition of energy flux is not intuitively clear. We could show that the proposed flux of potential energy neglects the energy of electron-hole excitations that is eventually given back to the edge \cite{stabler_transmission_2022}. This backaction effect is responsible for a perfectly quantized heat flux even in the presence of intrinsic dissipation. We showed this by introducing a transmission line model of the disorder at the edge; i.e. a Caldeira-Leggett type model where impurities are represented by periodic heat baths interacting with an edge state. More specifically we modeled a transmission line consisting of ohmic contacts connected via chiral edge states, see \cref{fig:TL}. The main advantage of this model is the unambiguous definition of the energy flux carried by the edge state, since in between the reservoirs the flux is described by free chiral bosonic edge states. The exact cancellation of the dissipated energy and the backaction effect is a direct consequence of the chirality of the edge channels. In the following we will introduce a two-node and transmission line model similar to our analysis in \cite{stabler_transmission_2022} with the difference that we introduce nonlocal interactions or chirality breaking diffusive currents between the reservoirs. Nonlocal refers to interactions with a range of the order of the disorder length scale  $\xi$. This breaks the chirality of the system locally and allows heat to be passed from one reservoir to the next directly without the help of a chiral edge state. The consequence of this is that the edge state carries an anomalously high heat flux, compensated by a deficit current passed by the nonlocal interactions or chirality breaking diffusive currents mediated by quantum point contacts.

\section{Backaction effect of Langevin Sources and its exact cancellation} \label{sec:back}

\begin{figure}[htbp]%
    \centering
    \subfloat[\centering Transmission line]{\includegraphics[width=.5\linewidth,valign=c]{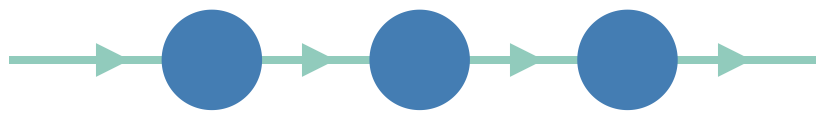}%
    \vphantom{\includegraphics[width=.45\linewidth,valign=c]{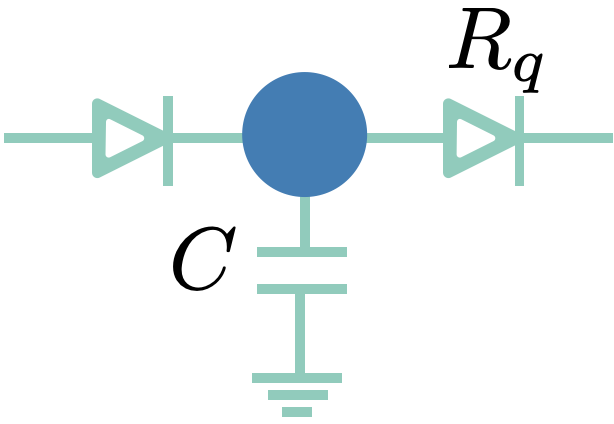}}
    \label{fig:TL}}%
    \hfil
    \subfloat[\centering Equivalent circuit]{{\includegraphics[width=.45\linewidth,valign=c]{reslocalTL.png} }\label{fig:effectivecircuit}}%
    \caption{The transmission line and the equivalent circuit of a single reservoir of the transmission line. The longitudinal current is chiral, indicated by the diodes, which have a quantum resistance $R_q= \frac{2\pi \hbar}{e^2}$.  Each node has a self capacitance $C$ and is assumed to be floating.}%
    \label{fig:TL+effectivecircuit}%
\end{figure}

Let us consider a QH edge state as a chiral, one dimensional channel. Each of the edge states carries heat proportional to the bosonic current-current correlation function \footnote{\cref{eq:heat} follows from writing a continuity equation for the Hamiltonian density $\hat{h}=\frac{\hbar v_F}{4\pi} \left(\partial_x \phi(x,t) \right)^2$, using the equation of motion and the definition of bosonic charge density $\rho(x,t)=\frac{e}{2\pi} \partial_x \phi(x,t)$ and bosonic current density $j(x,t)=-\frac{e}{2\pi} \partial_t \phi(x,t)$.}

\begin{equation}\label{eq:heat}
    J= \frac{R_q}{2} \left[\left\langle j^2(x,t) \right\rangle - \left\langle j^2(x,t) \right\rangle_{T=0} \right],
\end{equation} where $T$ indicates the temperature. For the free non-interacting edge $J$ evaluates to a universal value of a heat flux quantum $J_q= \frac{\pi k_B^2}{12 \hbar}T^2$.  In this paper, we will distinguish two different models to describe the edge defects, see \cref{fig:edgepic}. At first, we will model the edge defects by dissipative ohmic contacts, see also \cite{slobodeniuk_equilibration_2013}. In a second model, we will revert to the mesoscopic capacitor as a fully energy conserving approach using scattering theory. Under these assumptions we introduce dissipation in the system effectively by considering the edge states interacting with a lattice of heat baths, modeled as ohmic reservoirs \cite{slobodeniuk_equilibration_2013}, see \cref{fig:TL}. This can be understood as a variant of the Caldeira-Leggett model \cite{feynman_theory_1963,caldeira_path_1983}, where we assume a strong system bath coupling. A finite coupling between the system and bath will be considered separately.

The dynamics of the ohmic contacts are effectivley governed by three relevant energy scales; The level spacing of the reservoir $E_{\text{lvl}}$, the temperature of the reservoir $T$ and the charging energy $E_c$ due to the finite capacitance $C$ of the impurities. Physically a reservoir falls in either of the two categories; A small impurity has a large level spacing and resembles a two level system, however, if the impurity is large and hence the level spacing small, it resembles a many-level system. We will consider the limit of infinitely small level spacing. Finite level spacing effects are addressed later in the paper and the two level system limit will be discussed elsewhere. Furthermore we assume that we are in the Coulomb blockade regime and the following separation of energy scale holds $E_{\text{lvl}}\ll k_B T \ll E_c$.  

\begin{figure}[htbp]%
    \centering
    \includegraphics[width=.75\linewidth,valign=c]{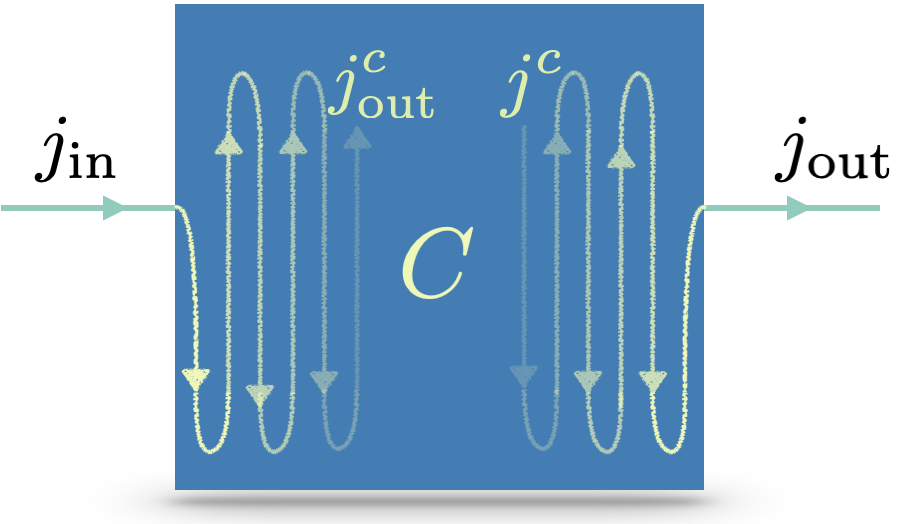}%
    \caption{Dynamics inside the reservoir. The capacitive interaction mediates scattering between the incoming current $j_{\text{in}}$, the Langevin source $j^c$, the current that is dissipated inside of the ohmic contact $j_{\text{out}}^c$ and the ourgoing current $j_{\text{out}}$. The scattering is energy conserving, which can be seen by the unitarity of the scattering matrix before tracing out the sources. }
    \label{fig:reservoir}%
\end{figure}

Each ohmic reservoirs is connected to the next one via a chiral edge state. The incident current creates electron-hole excitations, which  heat the reservoir and in turn leads to the emission of thermal (neutral) current fluctuations, alongside charge fluctuations, see \cref{fig:reservoir}. This allows us to use the Langevin equation approach. The equivalent circuit representation of each node is shown in \cref{fig:effectivecircuit}.

Each reservoir is characterized by a self-capacitance $C$, a longitudinal (quantum) resistance $R_q$. The corresponding Langevin equation tells us that at each node the outgoing (longitudinal) current emitted by the reservoir has two contributions

\begin{equation}
    j_{\text{out}}(t)=\frac{1}{\tau} Q(t)+j^c(t), \quad \tau=R_q C
\end{equation}

where all correlations between nodes are encoded in the collective mode $Q(t)$ and $j^c(t)$ is the Langevin source. This implies that the heat in the connecting edge state where the system is free and chiral consists of four contributions.

\begin{equation}
    J \sim \mathcal{C}_{cc}+ \frac{1}{\tau} \left(\mathcal{C}_{cq}+\mathcal{C}_{qc}\right)+\frac{1}{\tau^2}\mathcal{C}_{qq},
\end{equation} where we defined the source correlation function $\mathcal{C}_{cc}=\left\langle j^c(t) j^c(t) \right\rangle$, the source-charge correlation functions $\mathcal{C}_{cq} \sim \left\langle j^c(t)Q(t)\right\rangle$, $\mathcal{C}_{qc} \sim \left\langle Q(t)j^c(t)\right\rangle$ and the charge correlation function $\mathcal{C}_{qq} \sim \left\langle Q(t) Q(t)\right\rangle$. Note that the vacuum contribution at $T=0$ needs to be subtacted. 

We found the exact cancellation of the source-charge and charge-charge correlation functions upon integrating over the bandwidth of the system introduced by a Fourier transformation of the discrete lattice of reservoirs.

\begin{equation}
	\frac{1}{\tau}\int \! dk \ \mathcal{C}_{qq}(k,\omega)= -\!\!\int\! dk  \left( \mathcal{C}_{cq}(k,\omega) \!+\! \mathcal{C}_{qc}(k,\omega)\right),
\end{equation} 

This is one central result of \cite{stabler_transmission_2022}. The exact cancellation of correlation functions implies an energy balance; Heat absorbed by a reservoir $\sim \mathcal{C}_{qq}$ will later be reemitted as thermal fluctuaions $\sim \mathcal{C}_{cq}+ \mathcal{C}_{qc}$. This back action effect guarantees that the heat carried by the edge state is always an equilibrium heat flux quantum $J\sim \mathcal{C}_{cc}=J_q$  independent of the strength of dissipation introduced by the transverse interaction.

The result above can be understood in yet another way and is a direct consequence of the local nature of heat transport. Imagine cutting the transmission line into a segment, like the one shown in \cref{fig:TL}. No matter where the transmission line is cut between reservoirs, if the incoming current is equilibrium, the outgoing current is also going to be equilibrium due to energy conservation.

However nonlocal interactions, i.e. an interaction between nodes at different sites, or introducing a second nonchiral channel change this picture dramatically. Energy conservation only requires that the total longitudinal cross section carries an equilibrium amount of heat, but it is not clear how the energy will be distributed between the chiral channel and the second channel introduced by some nonlocal interaction or diffusion. Intuitively this already explains a potential enhancement or impediment of the heat flux in the chiral channel. To reveal this effect a local measurement of the correlation function in the chiral channel subject to the special interaction or diffusion is necessary. In this paper we will introduce two types of nonlocal interaction or diffusion between the nodes and their corresponding transmission lines and compute the heat carried by the edge state in the presence of these extra correlations of reservoirs.

\begin{comment}

\begin{figure}[htbp]
    \centering
    \includegraphics[width=\linewidth]{TL.png}
    \label{fig:TLCx_draft}
\end{figure}

\begin{figure}[htbp]
    \centering
    \includegraphics[width=\linewidth]{conservative.png}
    \label{fig:TLCx_draft}
\end{figure}

\begin{figure}[htbp]
    \centering
    \includegraphics[width=\linewidth]{conservativeCC.png}
    \label{fig:TLCx_draft}
\end{figure}

\begin{figure}[htbp]
    \centering
    \includegraphics[width=\linewidth]{resCC.png}
    \label{fig:TLCx_draft}
\end{figure}

\begin{figure}[htbp]
    \centering
    \includegraphics[width=\linewidth]{reslocalTL.png}
    \label{fig:TLCx_draft}
\end{figure}

\end{comment}

\section{Theoretical model - Nonlocal energy transport}

We will demonstrate that an edge channel can carry a nontrivial amount of heat in the presence of nonlocal interactions. The implications of this will be discussed later. We start from the simplest possible nontrivial model involving two ohmic reservoirs which are connected by an intermediate edge channel and are coupled via some non local interaction.

\subsection{Cross capacitive coupling}

\begin{figure}[htbp]%
    \centering
    \includegraphics[width=.9\linewidth,valign=c]{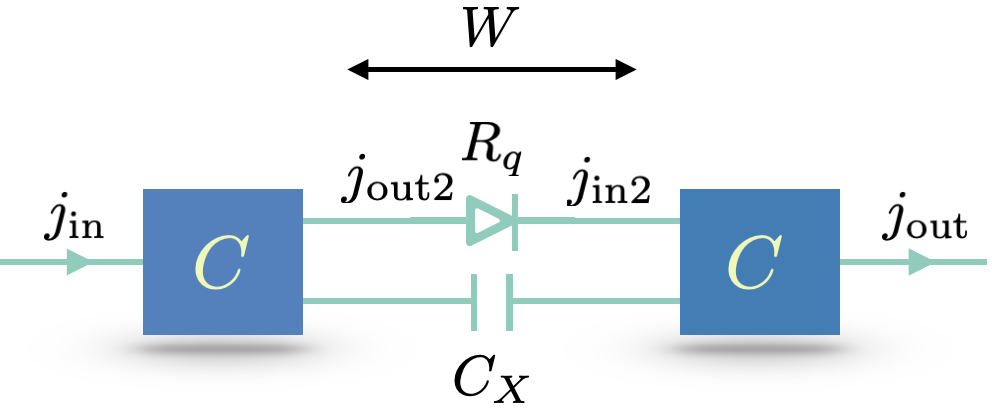}
    \caption{Cross capacitive coupling of reservoirs. The Hamiltonian density obtains additional terms of the form $C_X^{-1} Q_j Q_{j'}$ with $j\neq j'$, directly coupling the charge fluctuations of the reservoirs capacitively. One can formally solve the equations of motion inside of the reservoirs, i.e. of the edge states subject to the long-range Coulomb interaction, in terms of the boundary currents indicated in the figure. The equations of motions for the bosonic fields or currents are equivalent to the Langevin equations \cref{eq:CC2E1,eq:CC2E2,eq:CC2E3,eq:CC2E4}.}%
    \label{fig:resCCW}%
\end{figure}

The first example we would like to consider is  a cross capacitive coupling of strength $C_X$ between the charge fluctuations in each reservoir. This can be seen in \cref{fig:resCCW}.

  The starting point of our analysis is Kirchoff's law and the respective Langevin equations for the outgoing currents of the following form

 \begin{gather} \label{eq:CC2E1}
    \frac{d}{d t} Q_1(t) = j_{\text{in}}(t) - j_{\text{out}2}(t),\\\label{eq:CC2E2}
     \frac{d}{d t} Q_2(t) = j_{\text{in}2}(t) - j_{\text{out}}(t),\\\label{eq:CC2E3}
     j_{\text{out}2}(t)= \frac{1}{\tau}\left( Q_1(t)+\lambda  Q_2(t)\right) + j_2^{c}(t),\\\label{eq:CC2E4}
     j_{\text{out}}(t)= \frac{1}{\tau}\left( Q_2(t)+\lambda  Q_1(t)\right) + j^c(t),
\end{gather} with $\lambda=C/C_X \in \{0,1\}$ which can be solved for the outgoing currents $\Vec{j}_{\text{out}}=\left(j_\text{out},j_{\text{out}2},j^c_\text{out},j^c_{\text{out}2}\right)^T$ as a function of the incoming currents and sources $\Vec{j}_{\text{in}}=\left(j_\text{in},j_{\text{in}2},j^c,j_2^{c}\right)^T$. Note that the index $c$ refers to the neutral current propagating inside the ohmic contact. This gives the following relationship for the Fourier transformation of the currents in terms of a scattering matrix $\Vec{j}_{\text{out}}=\bm{\mathcal{M}} \cdot  \Vec{j}_{\text{in}}$

\begin{gather}
    \bm{\mathcal{M}} = \begin{pmatrix}
    \mathcal{A} &\mathcal{B}  & \mathcal{C} &-\!\mathcal{A}\\
    \mathcal{B} & \mathcal{A}  &  -\!\mathcal{A}& \mathcal{C}\\
    -\!\mathcal{A} & \mathcal{C} & \mathcal{B} & \mathcal{A}\\
    \mathcal{C}  & -\!\mathcal{A} & \mathcal{A} & \mathcal{B}
\end{pmatrix}, \\ 
\mathcal{A} = \frac{i \lambda  \tau  \omega }{\lambda ^2-(1-i \tau  \omega )^2},\\ \mathcal{B}=1-\mathcal{C},\\
\mathcal{C} = \frac{i \tau  \omega  (1-i \tau  \omega )}{\lambda ^2-(1-i \tau  \omega )^2}.
\end{gather}  One can easily check that this scattering matrix is unitary $\bm{\mathcal{M}\mathcal{M}}^\dagger = \mathbbm{1}$  reflecting the energy conserving nature of the scattering matrix before the Langevin sources are traced out. Note that this is the same scattering matrix as the one  obtained in \cite{idrisov_thermal_2022}. We implement the additional constraint that $j_{\text{in}2}= \exp\left(i \omega W /v_F\right) j_{\text{out}2}$ and solve for the outgoing currents as a function of the incoming boundary current $j_{\text{in}}$ and the sources $j^c_\text{out}$ and $j_2^{c}$, respectively. We give the resulting matrix $   \left(j_{\text{out}}, j_{\text{out}2}\right)^T
     = \bm{\mathcal{M}'} \left(j_{\text{in}},j^c, j_2^c\right)^T
    $ for $W\rightarrow0$ \footnote{ We remark that if $W\rightarrow\infty$ the current-current correlation function contains fast oscillations compared to all other relevant energy scales. Upon averaging over those fast oscillations the intermediate currents loose their correlations and thus the effect of nonlocal heat transport vanishes. All currents become equilibrium. For finite $W$ one observes modulations/resonances of the heat flux, which cannot change the sign of the corrections, but arise naturally since there is a modulation of the interference between the charge fluctuations in the respective reservoirs, mediated by the nonlocal interaction.}.

\begin{gather}
    \bm{\mathcal{M}'} = \begin{pmatrix}
    \mathcal{D} & \mathcal{F}+\mathcal{G}  & \mathcal{E} \\
    \mathcal{D}+\mathcal{E} & \mathcal{F}  &  \mathcal{G}\\
\end{pmatrix}\!,\\
\mathcal{D} = \frac{\lambda ^2-1+i \lambda  \tau  \omega }{\lambda ^2-i \lambda  \tau  \omega -(1-i \tau  \omega )^2}, \\
\mathcal{E} = \frac{i \tau  \omega -i \lambda  \tau  \omega }{\lambda ^2-i \lambda  \tau  \omega -(1-i \tau  \omega )^2},\\
\mathcal{F}=-\frac{i \lambda  \tau  \omega }{\lambda ^2-i \lambda  \tau  \omega -(1-i \tau  \omega )^2},\\
\mathcal{G}=\frac{i \tau  \omega  (1-i \tau  \omega )}{\lambda ^2-i \lambda  \tau  \omega -(1-i \tau  \omega )^2}.
\end{gather}  Note that $\bm{\mathcal{M}'}$ is not the scattering matrix of the system, since it contains the intermediate current $j_{\text{out}2}$ and is thus not unitary \footnote{The new $3\times3$ scattering matrix connecting the incoming state and sources to the outgoing state and the two internal states of the resistors remains unitary. This fixes the temperature of the reservoirs to be equilibrium, since from the unitarity of the scattering matrix immediately follows that $J^c = J^c_{\text{out}}$ and the same for the primed variables. This means that the Langevin source $j^c$, the current that is dissipated in the ohmic contact $j^c_{\text{out}}$ and the incoming current $j_{\text{in}}$ all have a noise power with the same equilibrium temperature.}. This means that part of the heat is passed by the nonlocal interaction.

Next, we assume that the incoming currents and Langevin sources are in equilibrium with the same temperature $\beta=\frac{1}{k_B T}$ {\color{blue}\footnote{Studying out of equilibrium situations like connecting the circuit or later the transmission line to a hot contact is an interesting and open question.  However, we want to address the situation after equilibration has taken place and postpone the nonequilibrium and steady state properties of the circuit to a later point.}}. This assumption is justified since the energy balance between absorbed and emitted heat $J_{\text{out}}^c=J^c$ is trivially fulfilled by the assumption of equal temperature for each reservoir. The current-current correlation function is given by
\begin{equation} \label{eq:specdens}
    \left\langle \delta j_x(\omega) \delta j_y(\omega') \right\rangle = 2\pi \delta\left(\omega+\omega'\right) \delta_{x,y} S(\omega),
\end{equation} $x,y \in \{\text{in},c,c2\}$ with the noise power $ S(\omega) = R_q^{-1} \hbar \omega \left(1-\exp\left(-\beta\hbar\omega \right)\right)^{-1}$. Using \cref{eq:heat} we are able to compute the heat flux carried by the edge states after and between the reservoirs.

\subsubsection{Heat flux carried by  {\textrm{\texorpdfstring{$j_{\emph{out}}$}{k} }}}

The unitarity of the scattering matrix guarantees energy conservation for the total system. This can be best seen from the outgoing current-current correlation function $\left\langle j_{\text{out}}(\omega)j_{\text{out}}(-\omega) \right\rangle = S(\omega) $, which translates to a full quantum of heat carried by the outgoing current. The same is true for the correlation functions of the internal states $\left\langle j^c_{\text{out}}(\omega)j^c_{\text{out}}(-\omega) \right\rangle$ and $\left\langle j^c_{\text{out}2}(\omega)j^c_{\text{out}2}(-\omega) \right\rangle$. Inserting  \cref{eq:specdens} in \cref{eq:heat} indeed reproduces the equilibrium result $J=J_q$.

\subsubsection{Heat flux carried by  {\textrm{\texorpdfstring{$j_{\emph{out}2}$}{k} }}}

 The result for $\left\langle j_{\text{out}2}(\omega)j_{\text{out}2}(-\omega) \right\rangle\neq S(\omega) $ is different. The edge state obtains additional correlations due to the nonlocal interactions. We find

\begin{gather}\label{eq:CCheat1res}
   \left\langle j_{\text{out}2}(\omega)j_{\text{out}2}(-\omega) \right\rangle  = f(\omega,\lambda) S(\omega), \\ \label{eq:CCheat2res}
   f(\omega,\lambda) \!=\!  \frac{\left(\lambda^2 \!- \!1 \right)^2 \!+\!\left(2\!+\!\lambda ^2\right) \tau^2 \omega ^2 \!+\!\tau^4 \omega ^4}{\left(\lambda ^2\!-\!1\right)^2\!+\!(2\!+\!\lambda  (3 \lambda\! -\!4 )) 
   \tau^2 \omega ^2 \!+\! \tau^4 \omega ^4 },
\end{gather} where $f(\omega,\lambda)$ is always larger than $1$ and for low temperatures $\tau k_B T /(\hbar) \ll 1, \left(1-\lambda ^2\right) $  can be approximated as $f(\omega,\lambda)-1\approx 2 \lambda (2-\lambda )   \left(\lambda ^2-1\right)^{-2} \tau^2 \omega ^2$. Note that this expansion breaks down if $\lambda \rightarrow 1$ is sufficiently close to the strongly interacting limit. If we look at the strongest possible interaction we find

\begin{equation}
    f(\omega,1) = \frac{3+\tau ^2 \omega ^2}{1+\tau ^2 \omega ^2} \approx 3 - 2\tau^2 \omega^2, 
\end{equation} signifying the emergence of a new regime with a different enhancement already starting in the constant order  $\mathcal{O}(1)$. One can estimate a critical temperature $T_c$ at a given $\lambda$ or vice versa by comparing the coefficients in the expansions at low temperatures and $\lambda=1$. This emergence of a strongly interacting regime is crucial to understanding the physics of the corresponding transmission line, where this enhancement  can be attributed to the emergence of a second ``soft" collective mode.

We  conclude that there is a negative heat drag in a system with a cross capacitive coupling.  This can be seen explicitly, by performing the integration necessary in \cref{eq:heat} using \cref{eq:CCheat1res} in the limit we have just demonstrated. In leading order the heat flux will be threefold larger than the equilibrium flux, which corresponds to an increase of the effective temperature by a factor of $T_{\text{eff}}\approx\sqrt{3}T_{0}$, where $T_{0}$ corresponds to the temperature of the injected equilibrium heat flux.   We conclude that the correction to the equilibrium heat flux for low temperatures appears as a $(k_B T)^4$ correction for weak and as $(k_B T)^2$ correction for strong interactions and has a positive sign, independently of $\lambda$. The enhancement stems from additional correlations due to out-of-phase fluctuations of the charge in the two different reservoirs resulting in a correlated state in the chiral channel.

\subsection{Cross capacitive coupling - Transmission line}
\label{sec:TLCC}
\begin{figure}[htbp]%
    \centering
    \includegraphics[width=.9\linewidth,valign=c]{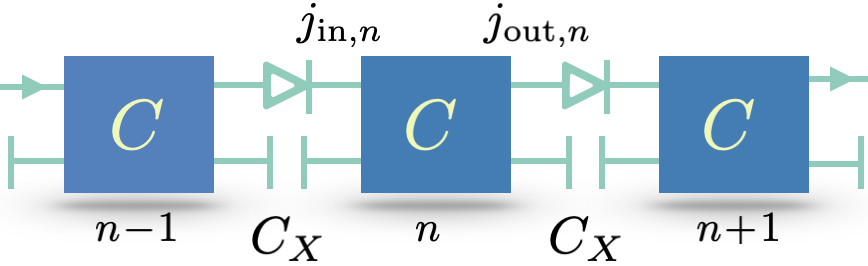}
    \caption{Transmission line with cross capacitive coupling. We extend our two node model by introducing nearest neighbor capacitive interaction between reservoirs. Similar to before the equations of motion can be either found from solving the Hamiltonian equations of motion directly or by solving the corresponding Langevin equations.}%
    \label{fig:TLCC}%
\end{figure}

We note that the transmission line (TL) model is not just a simple extension of the two reservoir model, but it contains some additional physics, namely the appearance of a second collective mode, which has important implications on the transport of heat. Let us consider a TL of reservoirs coupled via cross capactive interactions, see \cref{fig:TLCC}.

Every chiral edge channel is now accompanied by a second nonlocal channel in each crossection of the infinite line. The equations of motions for this case are given by

\begin{gather}\label{eq:TLCC1}
    \frac{d}{dt}Q_n(t)=j_{\text{in},n} -j_{\text{out},n},\\\label{eq:TLCC2}
    j_{\text{out},n} = \frac{1}{\tau} \mathcal{Q}_n(t) +  j_{\text{out},n}^c(t),\\\label{eq:TLCC3}
    \mathcal{Q}_n(t)= Q_n(t) + \lambda \left( Q_{n-1}(t) + Q_{n+1}(t) \right)
\end{gather} where the first equation implies charge conservation at each node and the second equation is a modified Langevin equation takeing into account the nonlocal interaction as a function of the dimensionless parameter $\lambda=C/C_X$. We promote these equations to transmission line equations, by demanding that $ j_{\text{in},n}(\omega)=j_{\text{out},n-1}(\omega)$. Since the dissipation leads to a decay of the collective mode, we will consider a large number of nodes and assume periodic boundary conditions, which allows us to use the following discrete Fourier transformations

\begin{gather} \label{eq:FT1}
	X_n(t)=\sum\limits_{k=0}^{N-1} \int \frac{d{\omega}}{2\pi} e^{i\frac{2 \pi k}{N} n-i\omega t} X_k(\omega),\\  \label{eq:FT2}	X_k(\omega)=\frac{1}{N}\sum\limits_{n=0}^{N-1} \int d{t} \  e^{-i \frac{2 \pi k}{N}  n+i \omega t} X_n(t),
\end{gather} and take the large N limit 
\begin{equation*}
    \lim\limits_{N \rightarrow \infty}\frac{1}{N}\sum_{k=0}^{N-1} \rightarrow \int_{-\pi/\xi}^{\pi/\xi} \frac{d{k}}{2\pi},
\end{equation*} where we introduced the distance between the nodes $\xi$. The general procedure to compute the heat flux carried by the edge state, in between reservoir $n$ and $n+1$,  involves the following steps:(1) Solve \cref{eq:TLCC1,eq:TLCC2,eq:TLCC3} using a Fourier transformation for $j_{\text{out}}(k,\omega)$. (2) Compute the correlation function $\mathcal{C}_{\text{inter}}(k,\omega)=\left\langle j_{\text{out}}(k,\omega)j_{\text{out}}(-k,-\omega) \right\rangle$ and integrate over frequencies and momenta. (3) Solve the momentum integral, by mapping it onto a unit circle contour with the transformation $z\rightarrow \exp\left(i k \xi \right)$ and pick up the residues inside of the contour. The outgoing current-current correlation function is given by

\begin{equation}
   \mathcal{C}_{\text{inter}}(k,\omega) = \frac{\omega^2 S(\omega)}{\left(\omega -  \varepsilon(k)\right)\left(\omega +  \varepsilon(-k)\right) },
\end{equation} where the spectrum of the collective mode is given by

\begin{equation}\label{eq:spec}
    \varepsilon(k)=\frac{i}{\tau} \left(e^{-i k \xi }-1\right) \left(1+  2 \lambda   \cos (k \xi )\right).
\end{equation} The eigenvalues of the capacitance matrix $C_{nm}= C^{-1} \left( \delta_{n,m} + \lambda \left( \delta_{n,m-1} + \delta_{n,m+1} \right) \right)$ \footnote{Before tracing out the sources, the system can be understood as a Hamiltonian system, see \cref{eq:Ham}, where different nodes interact according to a general capacitance matrix involving self- and cross-capacitances. This matrix is required to be positive definite by construction, which allows us to find the corresponding allowed value of $\lambda$.   } are given by $C^{-1}\left(1+  2 \lambda   \cos (k \xi )\right)$, which have to be positive definite. This limits $\lambda \in \{0,1/2\}$. The integral over momentum over the unit circle contour has the form

\begin{equation} \label{eq:ointTLCC}
 \Tilde{\mathcal{C}}_{\text{inter}}(\omega) =    \oint_\mathcal{C} \frac{ dz}{2\pi i} \frac{\omega^2 S(\omega)}{\left(z\omega -  z\Tilde{\varepsilon}(z)\right)\left(\omega +  \Tilde{\varepsilon}(1/z)\right) }, 
\end{equation} with $\Tilde{\varepsilon}(z)=\frac{i}{\tau} \left(\frac{1}{z}-1\right) \left(1+   \lambda \left(z+\frac{1}{z}\right)\right)$, where $ \Tilde{\mathcal{C}}_{\text{inter}}(\omega) =  \int \frac{dk}{2\pi}  \mathcal{C}_{\text{inter}}(k,\omega)  $ .  In general this integral has six poles with a complicated pole structure. We solve the integral exactly for the non interacting ($\lambda\rightarrow0$) and strongly interacting ($\lambda\rightarrow 1/2$) limit and compute low temperature deviations from that point and compute the crossover energy scale between the two limits. In the following we will analyze the heat flux $J=\frac{R_q}{4\pi} \int \mathop{d \omega} \Tilde{C}_{\text{inter}}(\omega)$ in the limits of weak and strong coupling.

\subsubsection{Non interacting limit \texorpdfstring{$\lambda\rightarrow 0$}{l0} }

In this limit the integral reduces to the one given in \cite{stabler_transmission_2022}. There is only one pole inside of the contour at $z_1=(1-i \tau \omega )^{-1}$ and one pole outside of the contour at $z_2=1+i\tau \omega$. We pick up $z_1$ and immediately find that $
 \Tilde{\mathcal{C}}_{\text{out}}(\omega) = 1, 
$ which gives a heat flux quantum, due to the local nature of the interaction and unitarity of the scattering matrix, as seen in \cite{stabler_transmission_2022}.

\subsubsection{Strongly interacting limit \texorpdfstring{$\lambda\rightarrow \frac{1}{2}$}{l12} }

In the opposite limit of strong interactions, the poles of  \cref{eq:ointTLCC} follow from the two cubic equations 
\begin{gather}
    1-z^2-z^3+z (1+2 i \tau \omega ) \overset{!}{=}0\\
    1+z-z^3-z^2 (1-2 i \tau \omega) ) \overset{!}{=}0.
\end{gather} We find that one pole is always outside of the contour, one pole is always inside of the contour, two poles are only inside for $\omega > 0$ and the other two poles are only inside for $\omega<0$. A detailed discussion of the pole structure was done in \cref{app:pole}. We pick up the poles inside of the contour and find that for low temperatures the correction to the heat flux quantum is given by

\begin{multline}\label{eq:heat12}
    J= \int_0^\infty \frac{d\omega}{2\pi} \left( 1+\frac{\sqrt{\tau \omega }}{2 \sqrt{2}}\right) \left(S(\omega)-S_{\text{vac}}(\omega)\right)\\
    = J_q + \frac{3}{8} \sqrt{\frac{\pi}{2}} \zeta\left(\frac{5}{2}\right) (k_B T)^{\frac{5}{2}} \sqrt{\frac{\tau}{\hbar^3}},
\end{multline} where $\zeta(s) = \sum\limits_{n=1}^\infty n^{-s}$ is the Riemann-Zeta function. 

\subsubsection{Crossover between the two limits}

It can be easily seen, e.g. by a perturbative expansion in small $\lambda$  that corrections to the $\lambda \approx 0$ regime will come with an additional factor of $\tau^2 \omega^2$ and thus the heat flux will obtain a correction that scales as $k_B^4 T^4$. However we can go beyond this perturbative expansion by assuming that $\tau k_B T /(\hbar) \ll 1, \left(1-2\lambda \right) $  is small and look for corrections proportional to $\tau^2 \omega^2$. This gives the following contribution to the heat flux, which as we note again is only pertrubative in temperature, but not in $\lambda$.

\begin{gather} \label{eq:heatpert}
   J=J_q + \frac{4 \pi ^4  \tau^2 k_B^4 T^4 }{15\hbar^3}  \mathcal{J}(\lambda),\\\label{eq:heatpert2}
  \mathcal{J}(\lambda)= \frac{\lambda}{(1+2 \lambda )^3}+\frac{\lambda(1-\lambda )}{(1-2 \lambda )^{3/2} (1+2 \lambda )^{5/2}}.
\end{gather}

 Upon analysing $\mathcal{J}(\lambda)$, we see that it diverges as $\lambda \rightarrow 1/2$. This signifies the breakdown of the perturbative expansion since the difference $1-2\lambda$ may become smaller as the energy scale set by temperature. This is the point where a second soft mode emerges. This can be seen from the pole structure of the propagator, the inverse Fourier transformation of $(-i\omega-\varepsilon(k))^{-1}$, with the spectrum from \cref{eq:spec}. In the strongly interacting limit there will be an additional pole at $\omega=0,\ k=\pi$ besides the pole at $\omega=0,\ k=0$. We can estimate the crossover from the single mode solution to the two mode solution by comparing the coefficients of  \cref{eq:heat12,eq:heatpert}. This gives the crossover value for the temperature $T_c$ at a given value of $\lambda$ or the crossover value for   $\lambda_c$ at a given temperature if the expression is inverted
 
 \begin{gather}
 \frac{\tau k_B T_c }{\hbar}=\frac{ \left(\frac{45}{2} \zeta \left(\frac{5}{2}\right)\right)^{2/3} }{8
   \pi ^{7/3}}\left(\frac{1}{\mathcal{J}(\lambda )}\right)^{2/3}.
 \end{gather}

Note that the nonlocal effect due to the cross capacitive coupling is a dynamical effect. In the limit where the collective mode is frozen out  ($\tau\rightarrow 0$) the heat flux vanishes and is blocked by the ohmic reservoirs. The nonlocal heat transport can thus be understood as out of phase fluctuations (``$k=\pi$") leading to an enhancement of heat in the edge state. Qualitatively the limits of strong interactions give different results for the TL and the two node model. This is reflected in the fact that for $\lambda_{2\text{node}} = 1$ the two node model can be written as a single node with charging energy $\frac{1}{2 C}(Q_1+Q_2)^2$, while it cannot be written like that for the transmission line at $\lambda_{\text{TL}}\rightarrow 1/2$. The Hamiltonian of the system contains only the product of neighboring charges $\frac{1}{2C}\sum_j \left(Q_j^2 + Q_j Q_{j+1}\right) \neq \frac{1}{2C} \left(\sum_j Q_j\right)^2 $.

\subsubsection{Work done by one reservoir on the neighboring reservoirs}

To find the part of heat passed by the nonlocal interaction imagine drawing a virtual box around a single node of the transmission line shown in \cref{fig:TLCC}. Let this box have length $L$ and choose it such that the bosonic current density $j(x,t)=-\frac{e}{2\pi}\partial_t \phi(x,t)$ at the boundaries correspond to $j(0,t)=  j_{\text{in},n}(t)$ and $j(L,t)=  j_{\text{out},n}(t)$. Next, write a continuity equation for the Hamiltonian density of the section of transmission line inside of the box $\partial_t\left\langle \hat{h}_n \right\rangle + \partial_x J_n=0$, where $n$ labels the node. Note that in the reservoir model we take $L\rightarrow \infty$ at the end. For a detailed discussion, see  \cref{app:mesCap}. The Hamiltonian reads

\begin{gather} \label{eq:Ham}
    \mathcal{H}_n=\frac{\hbar v_F}{4\pi}  \int_0^L d{x} \left( \partial_x \phi_{ n}(x,t) \right)^2  + \mathcal{H}_C,\\
    \mathcal{H}_C= \frac{Q^2_n(t)}{2 C} + \frac{Q_n(t) Q_{n-1}(t)}{C_X},
\end{gather} where $Q_j(t)=\frac{e}{2\pi}\int_{0}^L dx \ \partial_x \phi_n(x,t)$ is the integrated charge density of the reservoir. The total energy current density in the box $J_n=-\partial_t \left\langle \mathcal{H}_n \right\rangle=0$ is zero since the total energy is conserved, but we are able to express everything in terms of incoming and outgoing heat fluxes plus some additional term. 
\begin{multline}\label{eq:ebalance}
     J_n= \frac{R_q}{2} \left[\left\langle j^2(L,t) \right\rangle - \left\langle j^2(0,t) \right\rangle \right]\\+ \lambda \left\langle Q_j(t) \partial_t Q_{j-1}(t) - Q_j(t) \partial_t Q_{j+1}(t)\right\rangle.
\end{multline}  The first two terms correspond to the heat flux outgoing $J_{\text{out}}\sim  \left\langle j^2(L,t) \right\rangle $ and incident $J_{\text{in}}\sim  \left\langle j^2(0,t) \right\rangle$  to the reservoir. In the absence of the cross capacitive coupling these two terms cancel exactly, due to the cancellation of Joule heating and back action effect of the Langevin sources discussed in \cref{sec:back}. In the presence of cross capacitive interaction the two  additional terms in \cref{eq:ebalance} describe the work done by one reservoir on its neighbors and vice versa. For the two node model it is easy to see, that 

\begin{equation}
   -i \omega \lambda \left\langle Q_1(\omega) Q_{2}(-\omega) \right\rangle - h.c. \sim 1\!-\! f(\omega,\lambda),
\end{equation} which exactly compensates the extra heat in the chiral channel. Each cross section thus carries only one flux quantum of heat as expected from thermodynamic arguments. However a local measurement might reveal an anomalous amount of heat in the chiral channel despite a globally trivial heat flux.

Note that \cref{eq:Ham} is the same Hamiltonian as the one used in \cite{idrisov_thermal_2022}, where heat drag was studied in chiral systems. Our analysis thus holds also for this case and we can compute the heat passed by the cross capacitive coupling for all regimes considered in \cite{idrisov_thermal_2022}.

\subsection{Tunnel coupling} 

\begin{figure}[htbp]%
    \centering
    \includegraphics[width=.9\linewidth,valign=c]{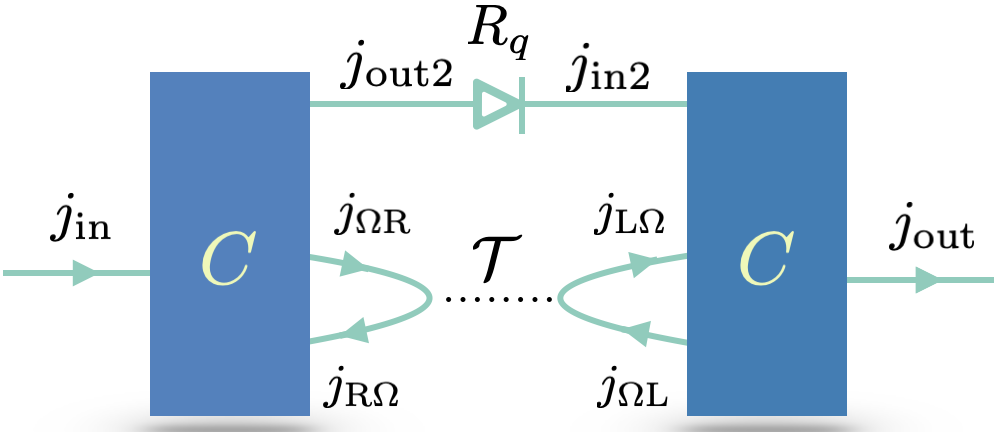}
    \caption{Direct tunnel coupling of the reservoirs. We consider QPCs in the regime where the transmission probability is energy independent. The tunneling of bosonic currents is in general non-unitary there, however the partitioning noise adjusts in a way that the energy balance between in- and outgoing heat at the QPC is respected.  }%
    \label{fig:2nodesQPC}%
\end{figure}

The second example of nonlocal interaction we would like to consider is a direct tunnel coupling of the reservoirs through a quantum point contact (QPC); see \cref{fig:2nodesQPC}.

The equation of motions are given by the following set of equation
\begin{gather}
    \frac{d}{d t} Q_1(t) =   j_{\text{in}}(t) - j_{\text{out}2}(t)  - j_{\Omega R}(t)+ j_{ R \Omega}(t),\\
    \frac{d}{d t} Q_2(t) =  j_{\text{in}2}(t) - j_{\text{out}}(t)-j_{\Omega  L}(t)+ j_{ L \Omega}(t),
    \end{gather} where the two equations are Kirchoff's law satisfying charge conservation of the two reservoirs. The two incoming currents $j_{L\Omega}$ and $j_{R\Omega}$, which passed the QPC are given by an energy independent scattering matrix ($\mathcal{T} \in \{0,1\}$) at the QPC \footnote{The transmission probability is only frequency independent if one considers the tunneling of free fermions. In leading order of the tunneling, the collective mode and source contribution are uncorrelated, which justfies this approach. Our conclusion holds for $ \mathcal{T}$  close to $0$ or $1$ and needs to be rechecked for arbitrary   $\mathcal{T}$ . This will be considered elsewhere.}
    \begin{gather}
    j_{ R \Omega}(t)= \mathcal{T} j_{\Omega L}(t)+(1-\mathcal{T}) j_{\Omega  R}(t)+j^Q(t),\\
    j_{ L  \Omega}(t)= \mathcal{T} j_{\Omega  R}(t)+(1-\mathcal{T}) j_{\Omega  L}(t)-j^Q(t),
    \end{gather}
    which is in general non-unitary, however the QPC introduces partioning nose $j^Q$, whose noise power is determined in a way such that the whole QPC is energy conserving and thus  $J_{ R \Omega}+J_{ L \Omega}=J_{  \Omega R}+J_{  \Omega L }$.

    Finally, all the currents outgoing from the reservoir are given by the Langevin equations respectively.
    \begin{gather}
    j_{\Omega  R} = \frac{1}{\tau}Q_1(t) + j^c_{  R}(t),\\
    j_{\Omega L} = \frac{1}{\tau}Q_2(t) + j^c_{ L}(t),\\
     j_{\text{out}2}(t)= \frac{1}{\tau}Q_1(t) + j_2^c(t),\\
     j_{\text{out}}(t)= \frac{1}{\tau} Q_2(t) + j^c(t).
\end{gather} We assume no retardation for the intermediate current $     j_{\text{in}2}(t)=j_{\text{out}2}(t)$ and solve the system of equations for the Fourier transformation of the currents. As before the temperatures of the reservoirs are equal to the temperature of the boundary current to the left, due to the unitarity of the scattering matrix inside of the reservoirs. Furthermore each crossection contains an equilibrium amount of heat, which can be seen by adding up the heat of the individual channels, e.g. $J_{\text{out}2}+J_{\Omega R}-J_{R \Omega} = J_q$ or $J_{\text{out}2}+J_{\Omega L}-J_{L \Omega} = J_q$, using \cref{eq:heat}. However each individual channel, between the reservoirs, carries a nontrivial amount of heat. The correlation function of the outgoing intermediate current is given by

\begin{gather}\label{eq:QPCheat1}
   \left\langle j_{\text{out}2}^2(\omega) \right\rangle  = \left(1+ g(\omega,\mathcal{T})\right) S(\omega), \\ \label{eq:QPCheat2}
   g(\omega,\mathcal{T}) =  \frac{2 \mathcal{T} \left(\Tilde{\mathcal{T}}- \omega ^2 \tau^2\right)}{\Tilde{\mathcal{T}}^2 + 2 \Tilde{\mathcal{T}} (\Tilde{\mathcal{T}}+ \mathcal{T})\omega ^2 \tau^2+ \omega ^4 \tau^4},
\end{gather} where $\Tilde{\mathcal{T}}=1+\mathcal{T}$. In contrast to the cross capacitive coupling we can safely perform a perturbative expansion for small frequencies $\tau k_B T /(\hbar) \ll 1 $ and find  

\begin{equation}
    g(\omega,\mathcal{T})\approx  \frac{2\mathcal{ T}}{1+\mathcal{T}} -\frac{2 \mathcal{T} (3+ 4\mathcal{T}) \tau ^2 \omega ^2}{(1+\mathcal{T})^2}. 
\end{equation} The correction thus starts already at the constant order and hence represents a direct correction to the value of the heat flux quantum. The heat carried by this channel is given by

\begin{equation}
    J_{\text{out}'} =  \left(1+\frac{2\mathcal{ T}}{1+\mathcal{T}}\right) J_q - \frac{2 \mathcal{T} (3+ 4\mathcal{T})   }{ (1+\mathcal{T})^2} J_4, 
\end{equation} where $J_4= \frac{\pi^4}{15 } \frac{\tau^2 }{ \hbar^3} \left(k_B T\right)^4$. There is a frequency independent part of the correction. The term $\frac{2\mathcal{ T}}{1+\mathcal{T}} J_q$ survives in the limit where all dynamics are suppressed $\tau \rightarrow 0$. In contrast to the cross capacitive coupling where the enhancement was dynamical, this represents a static contribution, a frequency independent neutral excitation of electron-hole pairs, which create additional correlations on the chiral channel.

\subsection{Tunnel coupling - Transmission line}

\begin{figure}[htbp]%
    \centering
    \includegraphics[width=\linewidth,valign=c]{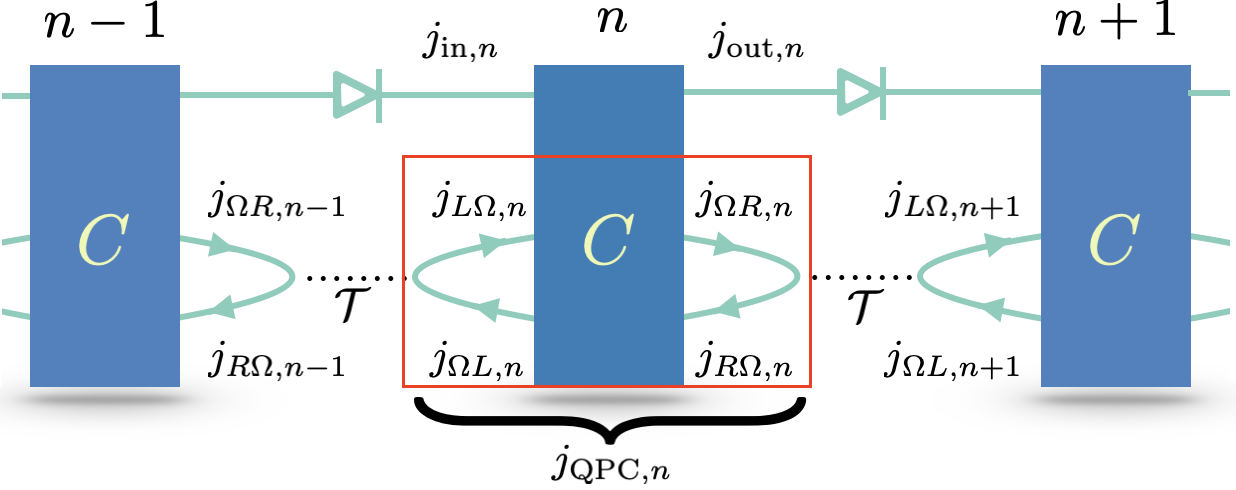}
    \caption{Transmission line with tunnel coupling. We extend our two node model by introducing nearest neighbor tunneling between reservoirs in the form of a QPC. Similar to before the equations of motion can be either found from solving the Hamiltonian equations of motion directly or by solving the corresponding Langevin equations.}%
    \label{fig:TLQPC}%
\end{figure}

Similar to before we promote the two reservoir model to a transmission line. The modified equation of motion have the following form 

\begin{equation}
    \frac{d}{d t} Q_n(t) =   j_{\text{in},n}(t) - j_{\text{out},n}(t)  + j_{\text{QPC},n},
    \end{equation} where $n$ labels the position of the node in the  line and $j_{\text{QPC},n}  \!=\! j_{ L \Omega,n}(t)- j_{\Omega L,n}(t) + j_{ R \Omega,n}(t)  - j_{\Omega R,n}(t),$ is the total net current of the incoming and outgoing from the left and right QPC; see \cref{fig:TLQPC}.
    
To obtain the transmission line equations we demand that $ j_{\text{in},n}(\omega)=j_{\text{out},n-1}(\omega)$. Since the dissipation leads to a decay of the collective mode, we will consider a large number of nodes and assume periodic boundary conditions, which allows us to use the discrete Fourier transformations \cref{eq:FT1,eq:FT2}. Solving for the heat flux follows the same steps already presented in \cref{sec:TLCC}.  The outgoing current $j_{\text{out},n}(t)$ statisfies a normal Langevin equation. The QPC current however statisfies a special  Langevin equation of the following form
 
 \begin{equation} \label{eq:jQPC}
     j_{\text{QPC}}(k,\omega) =  - \frac{2 \mathcal{T}}{\tau}  \left(1-\cos (k \xi )\right) Q(k,\omega) + j^s_{\text{QPC}},
 \end{equation} with a noisepower $S_{\text{QPC}}=(2R_q/R_{\text{QPC}}) S(\omega)$ with $R_{\text{QPC}}^{-1} =   2 \mathcal{T}R_q^{-1} \left(1-\cos (k \xi )\right) $. Note the difference of factor two due to non-chirality of the Langevin source. The sign in \cref{eq:jQPC} is due to the convention of the definition of current, however $R_{\text{QPC}}^{-1}$ can be understood as the Eigenvalues of the capacitance matrix of the system and are defined to be positive definite.   The spectrum of the collective mode  is given by 
 
 \begin{equation}\label{eq:specQPC}
    \varepsilon(k)=\frac{i}{\tau} \left(e^{-i k \xi }-1\right) - \frac{2i \mathcal{T}}{\tau}  \left(1-\cos (k \xi )\right).
\end{equation}

 The outgoing current-current correlation $\mathcal{C}_{\text{out},n}(k,\omega)=\left\langle j_{\text{out},n}(k,\omega)j_{\text{out},n}(-k,-\omega) \right\rangle$ function in the large-$N$ limit is given by

\begin{multline}
   \mathcal{C}_{\text{out},n}(k,\omega)= \frac{S_{\text{QPC}}(\omega)}{\tau ^2 (\omega +\epsilon (-k)) (\omega -\epsilon (k))}\\+ \frac{S(\omega) \left(1+e^{i k \xi } (i \tau  (-\omega \!+\!\epsilon (k))\!-\!1)\right) \left(\vphantom{e^{i k \xi }}\text{h.c.}\right)}{\tau ^2 (\omega +\epsilon (-k)) (\omega -\epsilon (k))}.
\end{multline} We map this onto the unit circle contour by doing the transformation $z\rightarrow e^{i k \xi}$ and pick up the poles inside of the contour. We find in total 6 poles, but in contrast to the capacitive coupling they stay inside or outside independently of $\omega$. For low temperatures we find the following correction to the heat flux.

\begin{equation}
\Tilde{ \mathcal{C}}_{\text{out},n}(\omega)\approx 1+\frac{2 \mathcal{T}}{1+\mathcal{T}}-6 \left(\mathcal{T} +2 \mathcal{T}^2 \right) \tau ^2 \omega ^2,
\end{equation} which is exact in $\mathcal{T}$, but perturbative in small frequencies. This gives a heat flux of

\begin{equation}
    J_{\text{out},n} =  \left(1+\frac{2\mathcal{ T}}{1+\mathcal{T}}\right) J_q -6 \left(\mathcal{T} +2 \mathcal{T}^2 \right) J_4,
\end{equation}where $J_4= \frac{\pi^4}{15 } \frac{\tau^2 }{ \hbar^3} \left(k_B T\right)^4$.  The heat flux is qualitatively similar to the one obtained for the two node case. Note also here the emergence of the direct enhancement of the heatflux quantum in the chiral channel. The heat carried by the QPC current is negative and exactly compensates the excess heat.

\section{Mesoscopic capacitors - energy conserving approach}
\label{app:mesCap}
\begin{figure}[htbp]%
    \centering
    \includegraphics[width=.45\linewidth,valign=c]{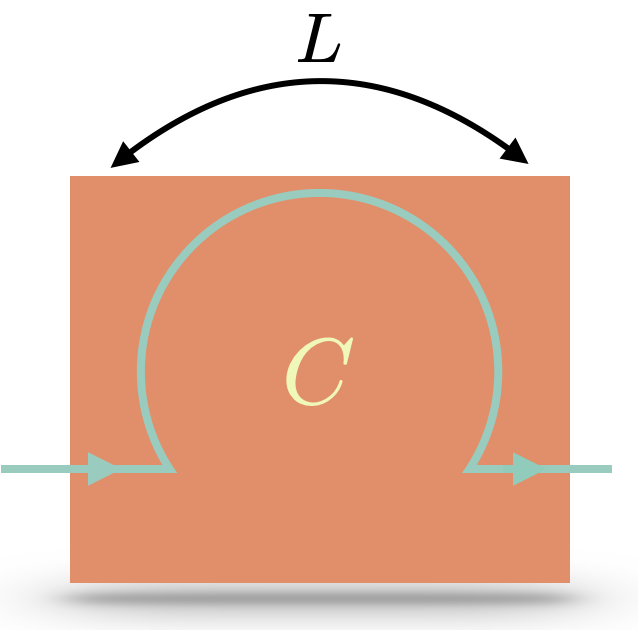}
    \caption{Mesocopic capacitor. A loop of edges state subject to Coulomb self interaction. The Hamiltonian inside of the loop aquires the additional interaction term $Q^2/2C$. Where Q is the total charge inside of the loop.}%
    \label{fig:mesCap}%
\end{figure}

One assumption we rely on is that all dissipative degrees of freedom lie in the ohmic reservoirs, which typically represent a metallic system with an infinitely small level spacing. The noise power of the Langevin sources can be represented  as an infinite number of harmonic oscillators.  This condition can be relaxed by replacing the ohmic reservoirs with mesoscopic capacitors \cite{roussel_electron_2017,litinski_interacting_2017}, a fully Hamiltonian and thus energy conserving system without dissipation. A mesoscopic capcitor is a loop of an edge state, which interacts via long range Coulomb interactions with itself, see \cref{fig:mesCap}.

\begin{figure}[htbp]%
    \centering
    \includegraphics[width=\linewidth,valign=c]{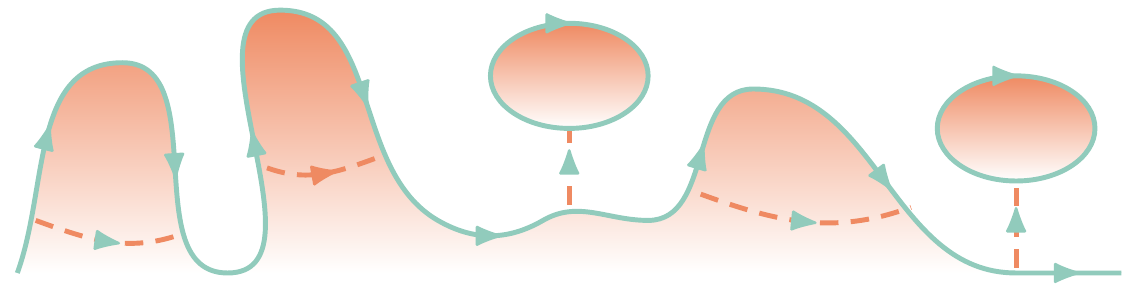}
    \caption{Defects in the edge.}%
    \label{fig:inhom}%
\end{figure}

The advantage for this is twofold. One one hand this mesoscopic capacitor is an easily realizable mesoscopic device; a building block to recreate our theoretical predictions in experiments. On the other hand this mesoscopic capacitor might be a more realistic approximation of the shape and form of disorder and defects inherent to QH edges, like shown in \cref{fig:inhom}. Our original motivation was to consider an effective model that includes dissipation intrinsically. However the assumption of modelling this as an ohmic reservoir remained an open question up until now. As we will show many of our results for the reservoirs qualitatively apply also to the energy conserving approach and might be an inherent property of QH edges.

In \cref{app: ConservativeCC} we  briefly revisit the cases of crosscapacitive coupling  with mesoscopic capacitors instead of ohmic reservoirs. Note the difference. The former does not rely on Langevin formalism, but on scattering theory and is a fully Hamiltonian system without any dissipation. As an example we considered two mesoscopic capacitors subject to cross capacitive interaction, but the special role of the mesoscopic capacitors can be directly generalized to tunnel coupling and the respective transmission lines. The main results are that a large capacitor $L \rightarrow \infty$ in equilibrium is equivalent to the reservoir model considered in this paper. The reason for this is that correlations between the currents propagating inside of the capacitor are lost, in a similar way as current is dissipated in the ohmic reservoir, which leads to the same physical situation and prediction for the correction of heat flux due to nonlocal interactions. Our analysis holds if the following separation of energy scale holds $\hbar v_F/L \ll k_B T \ll E_c$, where temperature $T$ is larger than the level spacing, but smaller than the charging energy of capacitor. 

For a finite length $L$ the result remains qualitatively the same as for the reservoirs, however the correlation function develops additional resonances due to interference effects between the charge fluctuations of the reservoirs.

\section{Conclusions}

We model chiral systems with effective instrinsic dissipation in the form of ohmic reservoirs. We rely on a combination of scattering theory and Langevin equations and have full access to equilibrium and non equilibrium situations. In thermal equilibrium, there is a cancellation between the Joule heating of the reservoirs and a back action effect of the Langevin sources. This cancellation is the reason for a robust quantization of heat carried by the edge state in the presence of dissipation and in the absence of chirality breaking nonlocal interactions or diffusion. 

We have shown that nonlocal interactions or diffusive currents between the reservoirs in a chiral system create highly correlated and nontrivial states, leading to a negative heat drag effect; an enhancement of the heat flux above the quantized value of the heat flux quantum $J_q= \frac{\pi k_B^2}{12 \hbar}T^2$ and thus a violation of the aforementioned energy balance in the chiral channel. This nonlocal passage of energy is in full agreement with the second law of thermodynamics, since every crossection of the system only carries an equilibrium amount of heat. This enhancement can be attributed to out of phase fluctuations for cross capacitively coupled reservoirs and to an enhancement through the Coulomb blockade effect in the case of a tunnel coupling. 

We have shown that neither dissipation nor  the specific choice of our model for a reservoir is the cause for this phenomenon. The same effect can be found by replacing the reservoirs with mesoscopic capacitors, a fully energy conserving system, consisting of a loop of the edge state subject to Coulomb self interaction. For a large mesoscopic capacitors the predictions are the same as for the reservoirs. For small mesoscopic capacitors the current correlation function develops resonances, but the effect remains qualitatively the same.

Our results show the non universality involved in heat transport experiments already in equilibrium and this paper can be understood as a toolbox to construct and detect these highly correlated and nontrivial states in mesoscopic experiments.

To proceed further one needs to look into the physics
of probing the correlated states. The states are highly correlated and thus in the tunneling Hamiltonian approach one needs to define an electronic operator which is not just the bosonized vertex
operator but goes beyond the bosonization formalism due to
correlations through the collective mode if a tunnel probe is
connected. It is an open question if one has experimental access to the charge of the excitations, e.g. by Aharonov Bohm type experiments \cite{idrisov_dephasing_2018}, to the quantum statistics e.g. by nonequilibrium measurements \cite{zhang_delta-t_2022} or to the correlation function directly by tunneling to a QD \cite{le_sueur_energy_2010}.

Finally, the formalism developed in this paper can be easily
applied to construct mesoscopic devices showing anomalous heat transport, unusual statistics and more interesting phenomena.

The authors acknowledges the financial support from
the Swiss National Science Foundation.

\appendix
\section{Evaluation of correlation functions - cross capacitive coupling}

\label{app:pole}

\subsection{Strongly interacting limit \texorpdfstring{$\lambda\rightarrow \frac{1}{2}$}{Al12} }

\begin{figure}[htbp]%
    \centering
    \subfloat[\centering Roots of $1-z^2-z^3+z (1+2 i \tau \omega ) \overset{!}{=}0$]{\includegraphics[width=.75\linewidth,valign=c]{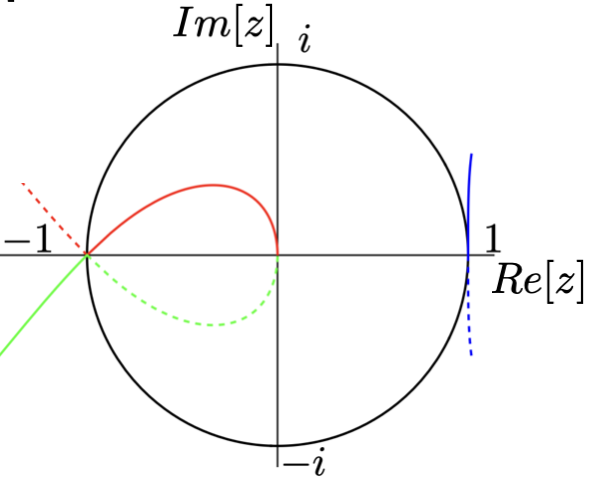}%
    \label{fig:p1}}%
    \hfil
    \subfloat[\centering  Roots of $1+z-z^3-z^2 (1-2 i \tau \omega) ) \overset{!}{=}0$]{{\includegraphics[width=.75\linewidth,valign=c]{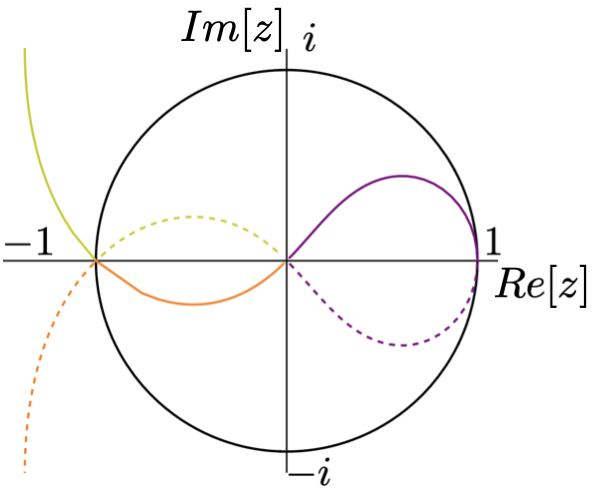} }\label{fig:p2}}%
    \caption{Poles of the heat integral in \cref{eq:ointTLCC}. The dashed lines indicate $\omega<0$ and solid lines $\omega>0$. The blue and purple poles are always present. In the non-interacting limit $\lambda\rightarrow0$ we only pickup the purple pole. The other poles around $ z=-1 \ \rightarrow \ k = \pi$ correspond to the second soft mode, the out-of-phase fluctuations of the collective mode. For finite $ \lambda $ these poles move towards the center of the contour, i.e. the momentum obtains a finite imaginary part. The excitation becomes gapped and we can rely on our low energy expansion, where the smallest energy scale is set by temperature, given by \cref{eq:heatpert}. }%
    \label{fig:p1+p2}%
\end{figure}

In this limit the poles of \cref{eq:ointTLCC} follow from the two polynomials 

\begin{gather}
    1-z^2-z^3+z (1+2 i \tau \omega ) \overset{!}{=}0\\
    1+z-z^3-z^2 (1-2 i \tau \omega) ) \overset{!}{=}0,
\end{gather} where $z \rightarrow e^{i k \xi}$. We refrain from writing the poles explicitly, but the structure of the poles can be seen in  \cref{fig:p1} for the first polynomial and \cref{fig:p2} for the second polynomial in  \cref{eq:ointTLCC}.

Note the structural difference of the poles in quadrant (I) and (IV), belonging to the collective mode contribution at $k \rightarrow 0 $ and the two poles in quadrant (II) and (III), which appear due to the nonlocal coupling and approach $k \rightarrow \pi $ for $\omega \rightarrow 0$ indicating the closing of a gap of a second mode, which becomes soft sufficiently close to this point. Upon picking up all three poles inside the contour for $\omega>0$ and $\omega<0$ seperatly and assuming small temperatures we find an intermediate result, which if integrated over frequencies directly yields \cref{eq:heat12}.

\subsection{Corrections to the regime \texorpdfstring{$\frac{\tau k_B T}{\hbar} \ll 1- 2 \lambda $}{AlC} }

If, however, temperature is the smallest energy scale in the system, this means the two additional poles are sufficiently far from the $k=\pi$ point, we can use the method of intermediate asymptotics, where we do a perturbative expansion in small temperature around the pole at $z\approx 1$ in the first and fourth quadrant, up to the necessary accuracy. This gives the first term of $\mathcal{J}(\lambda)$ in equation \cref{eq:heatpert2}. For the other two poles we can simply set $\omega\rightarrow 0$ in the denominator and pick up the other poles at  %
\begin{gather}
    z_{3/4}= -\frac{1}{2\lambda} + \sqrt{\frac{1}{4\lambda^2}-1}.
\end{gather} Since the whole correlation function is already proportional to $\tau^2\omega^2$ we immeadiatly find the correction from this pole by setting $\omega=0$ in the denominator. This gives the second term of $\mathcal{J}(\lambda)$ in equation \cref{eq:heatpert2}.

\section{Edge states with long range Coulomb cross coupling} \label{app: ConservativeCC}
\begin{figure}[htbp]%
    \centering
    \includegraphics[width=.9\linewidth,valign=c]{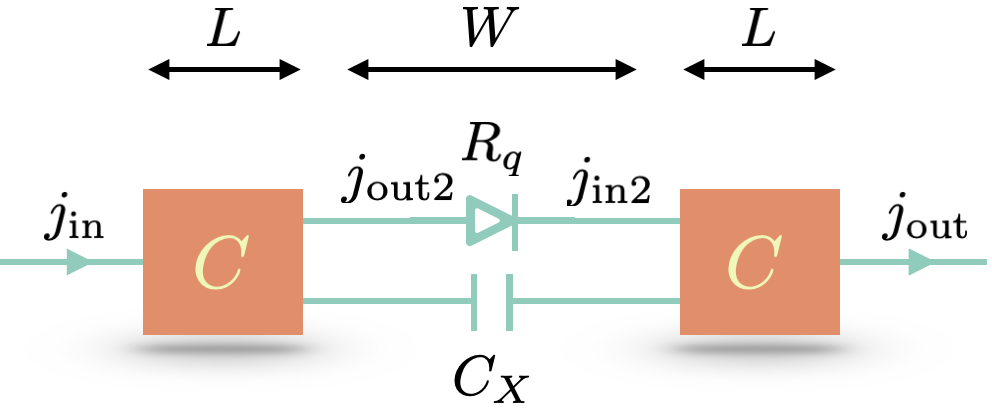}
    \caption{Cross capacitive coupling of mesoscopic capacitors.}%
    \label{fig:consCC}%
\end{figure}

 A loop of edge state of length $L$ interacts with itself via a long range Coulomb interaction with self capacitance $C$. We introduce a nonlocal Coulombic cross coupling $C_X$ between two of these mesoscopic capacitors which are separated by a distance $W$, see \cref{fig:consCC}. The Hamiltonian of the edge state inside of the interaction region acquires the following additional terms

\begin{multline}
    \mathcal{H}_j=\frac{\hbar v_F}{4\pi}  \int d{x} \left( \partial_x \phi_{ j}(x,t) \right)^2 \\+ \frac{Q^2_j(t)}{2 C} + \frac{Q_j(t) Q_{j-1}(t)}{C_X},
\end{multline} where $Q_j(t)=\frac{e}{2\pi}\int_0^L dx \ \partial_x \phi_j(x,t)$ is the integrated charge density of the loop and $j=1,2$ labels the left or right node.

The equation of motion of the bosonic field $\phi_j(x,t)$ is given by
\begin{equation}
     \partial_t \phi_{ j}(x,t) + v_F \partial_x \phi_{ j} (x,t) = - \frac{e}{\hbar C} \mathcal{Q}_j(t),  
\end{equation} with $\mathcal{Q}_j(t)= Q_j(t) + \lambda \left( Q_{j-1}(t) + Q_{j+1}(t) \right)$, where we introduce the relative coupling strength  $\lambda = \frac{C}{C_X} \in \{0,1\}$ and $Q_0(t)=Q_3(t)=0$. We assume the system is connected to a large reservoir without charge fluctuations and hence no capacitive cross couplings. The reservoir to the left has temperature $T$ and emits the equilibrium current $j_{\text{in}}$. We solve the equation of motion inside of the red interaction regions in terms of their boundary currents $j_{\text{in}},j_{\text{in}2},j_{\text{out}},j_{\text{out}2}$, which allows us to express all currents as a function of only the equilibrium current  $j_{\text{in}}$.  As expected we find that $\left\langle j_\text{out}^2(t) \right\rangle=\left\langle j_\text{in}^2(t) \right\rangle$. This is not necessarily the case for the intermediate current $j_{\text{out}2}$. The Fourier transformation of the outgoing current $\Vec{j}_{\text{out}}=\left(j_\text{out},j_{\text{out}2}\right)^T$ as a function of the incoming current $\Vec{j}_{\text{in}}=\left(j_\text{in},j_{\text{in}2}\right)^T$ is given by

\begin{gather}
    \Vec{j}_{\text{out}}=\mathbf{\mathcal{S}} \cdot  \Vec{j}_{\text{in}},\quad \mathbf{\mathcal{S}} = \begin{pmatrix}
\mathcal{A}(\omega) & \mathcal{B}(\omega) \\
\mathcal{B}(\omega) & \mathcal{A}(\omega)
\end{pmatrix},\\\label{eq:A}
\mathcal{A}\!=\!\frac{- i \lambda  \tau  \omega x^2(\frac{\omega}{2})}{ \left(1  \!-  \!\lambda
   ^2 \!-  \!i \tau  \omega \right) x^2(\frac{\omega}{2})\! +\!   \tau  \omega x(\omega) \!+  \!\tau ^2 \omega ^2 e^{-\frac{i\omega L }{v_F}}},\\\label{eq:B}
\mathcal{B}\!=\!\frac{ \left(1  \!-  \!\lambda ^2\right)  x^2(\frac{\omega}{2}) + \tau  \omega x(\omega)+\tau ^2 \omega ^2}{ \left(1  \!-  \!\lambda
   ^2 \!-  \!i \tau  \omega \right) x^2(\frac{\omega}{2})\! +\!   \tau  \omega x(\omega) \!+  \!\tau ^2 \omega ^2 e^{-\frac{i\omega L }{v_F}}},\\
   x(\omega)=2\sin \! \left(\frac{ \omega L }{v_F}\right),\quad \tau=R_q C.
\end{gather} The scattering matrix $\mathcal{S}$ is unitary, but there is another channel where heat can propagate. With the additional constraint that $j_{\text{in}2}=\exp(i \omega W /v_F)j_{\text{out}2}$, we can express the outgoing currents as a function of the boundary current $j_{\text{in}}$ only and compute their respective current- current correlation functions, i.e. the heat carried by the edge state. We analyse the non trivial correlation function $\left\langle j_{\text{out}2}^2(t) \right\rangle $ of the intermediate current in different limits of $W$ and $L$.

\subsubsection{Trivial limits}

We discover several trivial limits of the parameters $L$ and $W$. In the limit $L \rightarrow 0$ there is no accumulation of charge in the capacitor and hence no interaction between the reservoirs.

\begin{equation}
   \lim_{L\rightarrow 0}  \left\langle j_{\text{out}2}^2(t) \right\rangle=J_q .
\end{equation}

In the limit $W\rightarrow\infty$ the phase factors $\exp(\pm i \omega W /v_F)$ are fast oscillating and we average them over a period. We do this by mapping the average integration onto the unit circle contour $\mathcal{C}$ with the transformation $z\rightarrow\exp(i \omega W /v_F)$ and include only the residues inside the contour.

\begin{equation}\label{eq:Winfty}
 \left\langle j_{\text{out}2}^2(\omega) \right\rangle =    \oint_\mathcal{C} \frac{i dz}{2\pi} \frac{\left\langle j_{\text{in}}^2(\omega) \right\rangle|\mathcal{B}(\omega)|^2}{\left(z-\mathcal{A}(-\omega)\right)\left(z\mathcal{A}(\omega)-1\right)},
\end{equation} which evaluates to

\begin{equation}
\lim_{W\rightarrow \infty}  \left\langle j_{\text{out}2}^2(t) \right\rangle=J_q .
\end{equation}

 This means that the correlation between the outgoing and incoming intermediate currents is lost and the cross capacitive interaction does not influence the system and the current is equilibrium everywhere. For the rest of the paper we will thus consider the opposite limit of strong correlations between the intermediate currents and consider $W\rightarrow0$.

\subsubsection{Limit of $L\rightarrow \infty$ }

Similar to the case of $W\rightarrow\infty$ we average over the fast oscillations as a function of $L$ by the following transformation $z\rightarrow\exp(i \omega L /v_F)$. This gives a slightly different integral compared to \cref{eq:Winfty} of the form

\begin{equation}\label{eq:Linfty}
 \left\langle j_{\text{out}2}^2(\omega) \right\rangle \! =\!    \oint_\mathcal{C} \frac{i dz}{2\pi z} \frac{\left\langle j_{\text{in}}^2(\omega) \right\rangle \mathcal{B}_z(\omega)\mathcal{B}_{\frac{1}{z}}(-\omega)}{\left(1\!-\!\mathcal{A}_{\frac{1}{z}}(-\omega)\right)\! \!\left(\mathcal{A}_z(\omega)\!-\!1 \vphantom{\mathcal{A}_{\frac{1}{z}}(-\omega)}\right)},
\end{equation} where $\mathcal{A}_z(\omega)$ and $\mathcal{B}_z(\omega)$ is given by \cref{eq:A,eq:B} with the replacement $z\rightarrow\exp(i \omega L /v_F)$, note especially $x(\omega)=z+\frac{1}{z}$. $\mathcal{A}_\frac{1}{z}(-\omega)$ and $\mathcal{B}_\frac{1}{z}(-\omega)$ indicates the complex conjugate. The integral has three poles which are always inside the unit circle contour at 

\begin{align}
    z_0 &= 0,\\
    z_1 &= \frac{\lambda ^2-1+i \lambda  \tau  \omega }{(\lambda \! -\! 1) (1 \! +\! \lambda \! +\! i  \tau \omega  )\!+\!i \tau  \omega  \sqrt{\lambda  (\lambda +i \tau  \omega )}},\\
    z_2 &= \frac{(\lambda \! -\! 1) (1 \! +\! \lambda \! +\! i  \tau \omega  )\!+\!i \tau  \omega\sqrt{\lambda  (\lambda +i \tau  \omega )}}{\lambda ^2+i \lambda  \omega \tau -(1+i   \omega \tau)^2}.
\end{align}

Picking up the residues of this pole gives the non trivial result for the heat per frequency carried by the intermediate current

\begin{gather}\label{eq:CCheat1}
   \left\langle j_{\text{out}2}^2(\omega) \right\rangle  = f(\omega,\lambda) \left\langle j_{\text{in}}^2(\omega) \right\rangle, \\ \label{eq:CCheat2}
   f(\omega,\lambda) \!=\!  \frac{\left(\lambda \!- \!1 ^2\right)^2 \!+\!\left(2\!+\!\lambda ^2\right) \tau^2 \omega ^2 \!+\!\tau^4 \omega ^4}{\left(\lambda ^2\!-\!1\right)^2\!+\!(2\!+\!\lambda  (3 \lambda\! -\!4 )) 
   \tau^2 \omega ^2 \!+\! \tau^4 \omega ^4 },
\end{gather} where $f(\omega,\lambda)$ is the same correction as \cref{eq:CCheat2res}, the one we found with reservoirs instead of mesoscopic capacitors.

\subsubsection{Equivalence to the reservoir model}
 We note that the same equations \cref{eq:CCheat1,eq:CCheat2} can be found by replacing the large ($L\rightarrow\infty$) mesoscopic capacitors by ohmic reservoirs \cite{slobodeniuk_equilibration_2013}.  Note that the same corrections follows from solving the equations \cref{eq:CC2E1,eq:CC2E2,eq:CC2E3,eq:CC2E4} directly.

 The reason for this equivalence is the following. The current entering the mesoscopic capacitor propagating towards infinity looses the correlation with the current propagating from infinity towards the end of the capacitor. This means that an equilibrium source current, i.e. a Langevin source satisfies this condition.

\subsubsection{Finite $L$ or $W$}

In general finite $L$ or $W$ will introduce different kinds of oscillations of the heat carried by the intermediate edge state. This is natural and expected, since the negative drag effect arises from correlations between the charge in different nodes, which can be tuned by adjusting the retardation of the intermediate currents or the correlation of currents inside of the nodes themselves. This will lead to a modulation of the heat, but cannot change the sign of the correction to the heat flux.

\bibliographystyle{apsrev4-2}
%\bibliography{TL_Paper.bib}
\bibliography{Paper.bib}

\begin{thebibliography}{30}%
\makeatletter
\providecommand \@ifxundefined [1]{%
 \@ifx{#1\undefined}
}%
\providecommand \@ifnum [1]{%
 \ifnum #1\expandafter \@firstoftwo
 \else \expandafter \@secondoftwo
 \fi
}%
\providecommand \@ifx [1]{%
 \ifx #1\expandafter \@firstoftwo
 \else \expandafter \@secondoftwo
 \fi
}%
\providecommand \natexlab [1]{#1}%
\providecommand \enquote  [1]{``#1''}%
\providecommand \bibnamefont  [1]{#1}%
\providecommand \bibfnamefont [1]{#1}%
\providecommand \citenamefont [1]{#1}%
\providecommand \href@noop [0]{\@secondoftwo}%
\providecommand \href [0]{\begingroup \@sanitize@url \@href}%
\providecommand \@href[1]{\@@startlink{#1}\@@href}%
\providecommand \@@href[1]{\endgroup#1\@@endlink}%
\providecommand \@sanitize@url [0]{\catcode `\\12\catcode `\$12\catcode
  `\&12\catcode `\#12\catcode `\^12\catcode `\_12\catcode `\%12\relax}%
\providecommand \@@startlink[1]{}%
\providecommand \@@endlink[0]{}%
\providecommand \url  [0]{\begingroup\@sanitize@url \@url }%
\providecommand \@url [1]{\endgroup\@href {#1}{\urlprefix }}%
\providecommand \urlprefix  [0]{URL }%
\providecommand \Eprint [0]{\href }%
\providecommand \doibase [0]{https://doi.org/}%
\providecommand \selectlanguage [0]{\@gobble}%
\providecommand \bibinfo  [0]{\@secondoftwo}%
\providecommand \bibfield  [0]{\@secondoftwo}%
\providecommand \translation [1]{[#1]}%
\providecommand \BibitemOpen [0]{}%
\providecommand \bibitemStop [0]{}%
\providecommand \bibitemNoStop [0]{.\EOS\space}%
\providecommand \EOS [0]{\spacefactor3000\relax}%
\providecommand \BibitemShut  [1]{\csname bibitem#1\endcsname}%
\let\auto@bib@innerbib\@empty
%</preamble>
\bibitem [{\citenamefont {Kane}\ and\ \citenamefont
  {Fisher}(1997)}]{kane_quantized_1997}%
  \BibitemOpen
  \bibfield  {author} {\bibinfo {author} {\bibfnamefont {C.~L.}\ \bibnamefont
  {Kane}}\ and\ \bibinfo {author} {\bibfnamefont {M.~P.~A.}\ \bibnamefont
  {Fisher}},\ }\href {https://doi.org/10.1103/PhysRevB.55.15832} {\bibfield
  {journal} {\bibinfo  {journal} {Physical Review B}\ }\textbf {\bibinfo
  {volume} {55}},\ \bibinfo {pages} {15832} (\bibinfo {year}
  {1997})}\BibitemShut {NoStop}%
\bibitem [{\citenamefont {Banerjee}\ \emph {et~al.}(2018)\citenamefont
  {Banerjee}, \citenamefont {Heiblum}, \citenamefont {Umansky}, \citenamefont
  {Feldman}, \citenamefont {Oreg},\ and\ \citenamefont
  {Stern}}]{banerjee_observation_2018}%
  \BibitemOpen
  \bibfield  {author} {\bibinfo {author} {\bibfnamefont {M.}~\bibnamefont
  {Banerjee}}, \bibinfo {author} {\bibfnamefont {M.}~\bibnamefont {Heiblum}},
  \bibinfo {author} {\bibfnamefont {V.}~\bibnamefont {Umansky}}, \bibinfo
  {author} {\bibfnamefont {D.~E.}\ \bibnamefont {Feldman}}, \bibinfo {author}
  {\bibfnamefont {Y.}~\bibnamefont {Oreg}},\ and\ \bibinfo {author}
  {\bibfnamefont {A.}~\bibnamefont {Stern}},\ }\href
  {https://doi.org/10.1038/s41586-018-0184-1} {\bibfield  {journal} {\bibinfo
  {journal} {Nature}\ }\textbf {\bibinfo {volume} {559}},\ \bibinfo {pages}
  {205} (\bibinfo {year} {2018})}\BibitemShut {NoStop}%
\bibitem [{\citenamefont {Granger}\ \emph {et~al.}(2009)\citenamefont
  {Granger}, \citenamefont {Eisenstein},\ and\ \citenamefont
  {Reno}}]{granger_observation_2009}%
  \BibitemOpen
  \bibfield  {author} {\bibinfo {author} {\bibfnamefont {G.}~\bibnamefont
  {Granger}}, \bibinfo {author} {\bibfnamefont {J.~P.}\ \bibnamefont
  {Eisenstein}},\ and\ \bibinfo {author} {\bibfnamefont {J.~L.}\ \bibnamefont
  {Reno}},\ }\href {https://doi.org/10.1103/PhysRevLett.102.086803} {\bibfield
  {journal} {\bibinfo  {journal} {Physical Review Letters}\ }\textbf {\bibinfo
  {volume} {102}},\ \bibinfo {pages} {086803} (\bibinfo {year}
  {2009})}\BibitemShut {NoStop}%
\bibitem [{\citenamefont {le~Sueur}\ \emph {et~al.}(2010)\citenamefont
  {le~Sueur}, \citenamefont {Altimiras}, \citenamefont {Gennser}, \citenamefont
  {Cavanna}, \citenamefont {Mailly},\ and\ \citenamefont
  {Pierre}}]{le_sueur_energy_2010}%
  \BibitemOpen
  \bibfield  {author} {\bibinfo {author} {\bibfnamefont {H.}~\bibnamefont
  {le~Sueur}}, \bibinfo {author} {\bibfnamefont {C.}~\bibnamefont {Altimiras}},
  \bibinfo {author} {\bibfnamefont {U.}~\bibnamefont {Gennser}}, \bibinfo
  {author} {\bibfnamefont {A.}~\bibnamefont {Cavanna}}, \bibinfo {author}
  {\bibfnamefont {D.}~\bibnamefont {Mailly}},\ and\ \bibinfo {author}
  {\bibfnamefont {F.}~\bibnamefont {Pierre}},\ }\href
  {https://doi.org/10.1103/PhysRevLett.105.056803} {\bibfield  {journal}
  {\bibinfo  {journal} {Physical Review Letters}\ }\textbf {\bibinfo {volume}
  {105}},\ \bibinfo {pages} {056803} (\bibinfo {year} {2010})}\BibitemShut
  {NoStop}%
\bibitem [{\citenamefont {Venkatachalam}\ \emph {et~al.}(2012)\citenamefont
  {Venkatachalam}, \citenamefont {Hart}, \citenamefont {Pfeiffer},
  \citenamefont {West},\ and\ \citenamefont
  {Yacoby}}]{venkatachalam_local_2012}%
  \BibitemOpen
  \bibfield  {author} {\bibinfo {author} {\bibfnamefont {V.}~\bibnamefont
  {Venkatachalam}}, \bibinfo {author} {\bibfnamefont {S.}~\bibnamefont {Hart}},
  \bibinfo {author} {\bibfnamefont {L.}~\bibnamefont {Pfeiffer}}, \bibinfo
  {author} {\bibfnamefont {K.}~\bibnamefont {West}},\ and\ \bibinfo {author}
  {\bibfnamefont {A.}~\bibnamefont {Yacoby}},\ }\href
  {https://doi.org/10.1038/nphys2384} {\bibfield  {journal} {\bibinfo
  {journal} {Nature Physics}\ }\textbf {\bibinfo {volume} {8}},\ \bibinfo
  {pages} {676} (\bibinfo {year} {2012})}\BibitemShut {NoStop}%
\bibitem [{\citenamefont {Sivre}\ \emph {et~al.}(2019)\citenamefont {Sivre},
  \citenamefont {Duprez}, \citenamefont {Anthore}, \citenamefont {Aassime},
  \citenamefont {Parmentier}, \citenamefont {Cavanna}, \citenamefont {Ouerghi},
  \citenamefont {Gennser},\ and\ \citenamefont
  {Pierre}}]{sivre_electronic_2019}%
  \BibitemOpen
  \bibfield  {author} {\bibinfo {author} {\bibfnamefont {E.}~\bibnamefont
  {Sivre}}, \bibinfo {author} {\bibfnamefont {H.}~\bibnamefont {Duprez}},
  \bibinfo {author} {\bibfnamefont {A.}~\bibnamefont {Anthore}}, \bibinfo
  {author} {\bibfnamefont {A.}~\bibnamefont {Aassime}}, \bibinfo {author}
  {\bibfnamefont {F.~D.}\ \bibnamefont {Parmentier}}, \bibinfo {author}
  {\bibfnamefont {A.}~\bibnamefont {Cavanna}}, \bibinfo {author} {\bibfnamefont
  {A.}~\bibnamefont {Ouerghi}}, \bibinfo {author} {\bibfnamefont
  {U.}~\bibnamefont {Gennser}},\ and\ \bibinfo {author} {\bibfnamefont
  {F.}~\bibnamefont {Pierre}},\ }\href
  {https://doi.org/10.1038/s41467-019-13566-8} {\bibfield  {journal} {\bibinfo
  {journal} {Nature Communications}\ }\textbf {\bibinfo {volume} {10}},\
  \bibinfo {pages} {5638} (\bibinfo {year} {2019})}\BibitemShut {NoStop}%
\bibitem [{\citenamefont {Duprez}\ \emph {et~al.}(2021)\citenamefont {Duprez},
  \citenamefont {Pierre}, \citenamefont {Sivre}, \citenamefont {Aassime},
  \citenamefont {Parmentier}, \citenamefont {Cavanna}, \citenamefont {Ouerghi},
  \citenamefont {Gennser}, \citenamefont {Safi}, \citenamefont {Mora},\ and\
  \citenamefont {Anthore}}]{duprez_dynamical_2021}%
  \BibitemOpen
  \bibfield  {author} {\bibinfo {author} {\bibfnamefont {H.}~\bibnamefont
  {Duprez}}, \bibinfo {author} {\bibfnamefont {F.}~\bibnamefont {Pierre}},
  \bibinfo {author} {\bibfnamefont {E.}~\bibnamefont {Sivre}}, \bibinfo
  {author} {\bibfnamefont {A.}~\bibnamefont {Aassime}}, \bibinfo {author}
  {\bibfnamefont {F.~D.}\ \bibnamefont {Parmentier}}, \bibinfo {author}
  {\bibfnamefont {A.}~\bibnamefont {Cavanna}}, \bibinfo {author} {\bibfnamefont
  {A.}~\bibnamefont {Ouerghi}}, \bibinfo {author} {\bibfnamefont
  {U.}~\bibnamefont {Gennser}}, \bibinfo {author} {\bibfnamefont
  {I.}~\bibnamefont {Safi}}, \bibinfo {author} {\bibfnamefont {C.}~\bibnamefont
  {Mora}},\ and\ \bibinfo {author} {\bibfnamefont {A.}~\bibnamefont
  {Anthore}},\ }\href {https://doi.org/10.1103/PhysRevResearch.3.023122}
  {\bibfield  {journal} {\bibinfo  {journal} {Physical Review Research}\
  }\textbf {\bibinfo {volume} {3}},\ \bibinfo {pages} {023122} (\bibinfo {year}
  {2021})}\BibitemShut {NoStop}%
\bibitem [{\citenamefont {Goremykina}\ \emph {et~al.}(2019)\citenamefont
  {Goremykina}, \citenamefont {Borin},\ and\ \citenamefont
  {Sukhorukov}}]{goremykina_heat_2019}%
  \BibitemOpen
  \bibfield  {author} {\bibinfo {author} {\bibfnamefont {A.}~\bibnamefont
  {Goremykina}}, \bibinfo {author} {\bibfnamefont {A.}~\bibnamefont {Borin}},\
  and\ \bibinfo {author} {\bibfnamefont {E.}~\bibnamefont {Sukhorukov}},\
  }\href {http://arxiv.org/abs/1908.01213} {\bibfield  {journal} {\bibinfo
  {journal} {arXiv:1908.01213 [cond-mat]}\ } (\bibinfo {year}
  {2019})}\BibitemShut {NoStop}%
\bibitem [{\citenamefont {St\"abler}\ and\ \citenamefont
  {Sukhorukov}(2022)}]{stabler_transmission_2022}%
  \BibitemOpen
  \bibfield  {author} {\bibinfo {author} {\bibfnamefont {F.}~\bibnamefont
  {St\"abler}}\ and\ \bibinfo {author} {\bibfnamefont {E.}~\bibnamefont
  {Sukhorukov}},\ }\href {https://doi.org/10.1103/PhysRevB.105.235417}
  {\bibfield  {journal} {\bibinfo  {journal} {Physical Review B}\ }\textbf
  {\bibinfo {volume} {105}},\ \bibinfo {pages} {235417} (\bibinfo {year}
  {2022})}\BibitemShut {NoStop}%
\bibitem [{\citenamefont {Narozhny}\ and\ \citenamefont
  {Levchenko}(2016)}]{narozhny_coulomb_2016}%
  \BibitemOpen
  \bibfield  {author} {\bibinfo {author} {\bibfnamefont {B.}~\bibnamefont
  {Narozhny}}\ and\ \bibinfo {author} {\bibfnamefont {A.}~\bibnamefont
  {Levchenko}},\ }\href {https://doi.org/10.1103/RevModPhys.88.025003}
  {\bibfield  {journal} {\bibinfo  {journal} {Rev. Mod. Phys.}\ }\textbf
  {\bibinfo {volume} {88}},\ \bibinfo {pages} {025003} (\bibinfo {year}
  {2016})}\BibitemShut {NoStop}%
\bibitem [{\citenamefont {Levchenko}\ and\ \citenamefont
  {Kamenev}(2008)}]{levchenko_coulomb_2008}%
  \BibitemOpen
  \bibfield  {author} {\bibinfo {author} {\bibfnamefont {A.}~\bibnamefont
  {Levchenko}}\ and\ \bibinfo {author} {\bibfnamefont {A.}~\bibnamefont
  {Kamenev}},\ }\href {https://doi.org/10.1103/PhysRevLett.101.216806}
  {\bibfield  {journal} {\bibinfo  {journal} {Phys. Rev. Lett.}\ }\textbf
  {\bibinfo {volume} {101}},\ \bibinfo {pages} {216806} (\bibinfo {year}
  {2008})}\BibitemShut {NoStop}%
\bibitem [{\citenamefont {Raichev}\ \emph {et~al.}(2020)\citenamefont
  {Raichev}, \citenamefont {Gusev}, \citenamefont {Hernandez}, \citenamefont
  {Levin},\ and\ \citenamefont {Bakarov}}]{raichev_phonon_2020}%
  \BibitemOpen
  \bibfield  {author} {\bibinfo {author} {\bibfnamefont {O.~E.}\ \bibnamefont
  {Raichev}}, \bibinfo {author} {\bibfnamefont {G.~M.}\ \bibnamefont {Gusev}},
  \bibinfo {author} {\bibfnamefont {F.~G.~G.}\ \bibnamefont {Hernandez}},
  \bibinfo {author} {\bibfnamefont {A.~D.}\ \bibnamefont {Levin}},\ and\
  \bibinfo {author} {\bibfnamefont {A.~K.}\ \bibnamefont {Bakarov}},\ }\href
  {https://doi.org/10.1103/PhysRevB.102.195301} {\bibfield  {journal} {\bibinfo
   {journal} {Phys. Rev. B}\ }\textbf {\bibinfo {volume} {102}},\ \bibinfo
  {pages} {195301} (\bibinfo {year} {2020})}\BibitemShut {NoStop}%
\bibitem [{\citenamefont {Strait}\ \emph {et~al.}()\citenamefont {Strait},
  \citenamefont {Holland}, \citenamefont {Zhu}, \citenamefont {Zhang},
  \citenamefont {Ilic}, \citenamefont {Agrawal}, \citenamefont {Pacifici},\
  and\ \citenamefont {Lezec}}]{strait_revisiting_2019}%
  \BibitemOpen
  \bibfield  {author} {\bibinfo {author} {\bibfnamefont {J.~H.}\ \bibnamefont
  {Strait}}, \bibinfo {author} {\bibfnamefont {G.}~\bibnamefont {Holland}},
  \bibinfo {author} {\bibfnamefont {W.}~\bibnamefont {Zhu}}, \bibinfo {author}
  {\bibfnamefont {C.}~\bibnamefont {Zhang}}, \bibinfo {author} {\bibfnamefont
  {B.~R.}\ \bibnamefont {Ilic}}, \bibinfo {author} {\bibfnamefont
  {A.}~\bibnamefont {Agrawal}}, \bibinfo {author} {\bibfnamefont
  {D.}~\bibnamefont {Pacifici}},\ and\ \bibinfo {author} {\bibfnamefont
  {H.~J.}\ \bibnamefont {Lezec}},\ }\href
  {https://doi.org/10.1103/PhysRevLett.123.053903} {\bibfield  {journal}
  {\bibinfo  {journal} {Phys. Rev. Lett.}\ }\textbf {\bibinfo {volume} {123}},\
  \bibinfo {pages} {053903}}\BibitemShut {NoStop}%
\bibitem [{\citenamefont {Gurevich}\ and\ \citenamefont
  {Muradov}(2015)}]{gurevich_drag_2015}%
  \BibitemOpen
  \bibfield  {author} {\bibinfo {author} {\bibfnamefont {V.~L.}\ \bibnamefont
  {Gurevich}}\ and\ \bibinfo {author} {\bibfnamefont {M.~I.}\ \bibnamefont
  {Muradov}},\ }\href {https://doi.org/10.1134/S1063776115130026} {\bibfield
  {journal} {\bibinfo  {journal} {J. Exp. Theor. Phys.}\ }\textbf {\bibinfo
  {volume} {121}},\ \bibinfo {pages} {998} (\bibinfo {year}
  {2015})}\BibitemShut {NoStop}%
\bibitem [{\citenamefont {Idrisov}\ \emph {et~al.}(2022)\citenamefont
  {Idrisov}, \citenamefont {Levkivskyi},\ and\ \citenamefont
  {Sukhorukov}}]{idrisov_thermal_2022}%
  \BibitemOpen
  \bibfield  {author} {\bibinfo {author} {\bibfnamefont {E.~G.}\ \bibnamefont
  {Idrisov}}, \bibinfo {author} {\bibfnamefont {I.~P.}\ \bibnamefont
  {Levkivskyi}},\ and\ \bibinfo {author} {\bibfnamefont {E.~V.}\ \bibnamefont
  {Sukhorukov}},\ }\href {http://arxiv.org/abs/2203.02558} {\bibfield
  {journal} {\bibinfo  {journal} {arXiv:2203.02558 [cond-mat]}\ } (\bibinfo
  {year} {2022})}\BibitemShut {NoStop}%
\bibitem [{\citenamefont {Filliger}\ and\ \citenamefont
  {Reimann}(2007)}]{filliger_brownian_2007}%
  \BibitemOpen
  \bibfield  {author} {\bibinfo {author} {\bibfnamefont {R.}~\bibnamefont
  {Filliger}}\ and\ \bibinfo {author} {\bibfnamefont {P.}~\bibnamefont
  {Reimann}},\ }\href {https://doi.org/10.1103/PhysRevLett.99.230602}
  {\bibfield  {journal} {\bibinfo  {journal} {Phys. Rev. Lett.}\ }\textbf
  {\bibinfo {volume} {99}},\ \bibinfo {pages} {230602} (\bibinfo {year}
  {2007})}\BibitemShut {NoStop}%
\bibitem [{\citenamefont {Chiang}\ \emph {et~al.}(2017)\citenamefont {Chiang},
  \citenamefont {Lee}, \citenamefont {Lai},\ and\ \citenamefont
  {Chen}}]{chiang_electrical_2017}%
  \BibitemOpen
  \bibfield  {author} {\bibinfo {author} {\bibfnamefont {K.-H.}\ \bibnamefont
  {Chiang}}, \bibinfo {author} {\bibfnamefont {C.-L.}\ \bibnamefont {Lee}},
  \bibinfo {author} {\bibfnamefont {P.-Y.}\ \bibnamefont {Lai}},\ and\ \bibinfo
  {author} {\bibfnamefont {Y.-F.}\ \bibnamefont {Chen}},\ }\href
  {https://doi.org/10.1103/PhysRevE.96.032123} {\bibfield  {journal} {\bibinfo
  {journal} {Phys. Rev. E}\ }\textbf {\bibinfo {volume} {96}},\ \bibinfo
  {pages} {032123} (\bibinfo {year} {2017})}\BibitemShut {NoStop}%
\bibitem [{Note1()}]{Note1}%
  \BibitemOpen
  \bibinfo {note} {\protect \cref {eq:heat} follows from writing a continuity
  equation for the Hamiltonian density $\protect \hat {h}=\protect \frac
  {\protect \hbar v_F}{4\pi } \left (\partial _x \phi (x,t) \right )^2$, using
  the equation of motion and the definition of bosonic charge density $\rho
  (x,t)=\protect \frac {e}{2\pi } \partial _x \phi (x,t)$ and bosonic current
  density $j(x,t)=-\protect \frac {e}{2\pi } \partial _t \phi
  (x,t)$.}\BibitemShut {Stop}%
\bibitem [{\citenamefont {Slobodeniuk}\ \emph {et~al.}(2013)\citenamefont
  {Slobodeniuk}, \citenamefont {Levkivskyi},\ and\ \citenamefont
  {Sukhorukov}}]{slobodeniuk_equilibration_2013}%
  \BibitemOpen
  \bibfield  {author} {\bibinfo {author} {\bibfnamefont {A.~O.}\ \bibnamefont
  {Slobodeniuk}}, \bibinfo {author} {\bibfnamefont {I.~P.}\ \bibnamefont
  {Levkivskyi}},\ and\ \bibinfo {author} {\bibfnamefont {E.~V.}\ \bibnamefont
  {Sukhorukov}},\ }\href {https://doi.org/10.1103/PhysRevB.88.165307}
  {\bibfield  {journal} {\bibinfo  {journal} {Physical Review B}\ }\textbf
  {\bibinfo {volume} {88}},\ \bibinfo {pages} {165307} (\bibinfo {year}
  {2013})}\BibitemShut {NoStop}%
\bibitem [{\citenamefont {Feynman}\ and\ \citenamefont
  {Vernon}(1963)}]{feynman_theory_1963}%
  \BibitemOpen
  \bibfield  {author} {\bibinfo {author} {\bibfnamefont {R.~P.}\ \bibnamefont
  {Feynman}}\ and\ \bibinfo {author} {\bibfnamefont {F.~L.}\ \bibnamefont
  {Vernon}},\ }\href {https://doi.org/10.1016/0003-4916(63)90068-X} {\bibfield
  {journal} {\bibinfo  {journal} {Annals of Physics}\ }\textbf {\bibinfo
  {volume} {24}},\ \bibinfo {pages} {118} (\bibinfo {year} {1963})}\BibitemShut
  {NoStop}%
\bibitem [{\citenamefont {Caldeira}\ and\ \citenamefont
  {Leggett}(1983)}]{caldeira_path_1983}%
  \BibitemOpen
  \bibfield  {author} {\bibinfo {author} {\bibfnamefont {A.~O.}\ \bibnamefont
  {Caldeira}}\ and\ \bibinfo {author} {\bibfnamefont {A.~J.}\ \bibnamefont
  {Leggett}},\ }\href {https://doi.org/10.1016/0378-4371(83)90013-4} {\bibfield
   {journal} {\bibinfo  {journal} {Physica A}\ }\textbf {\bibinfo {volume}
  {121}},\ \bibinfo {pages} {587} (\bibinfo {year} {1983})}\BibitemShut
  {NoStop}%
\bibitem [{Note2()}]{Note2}%
  \BibitemOpen
  \bibinfo {note} {We remark that if $W\rightarrow \infty $ the current-current
  correlation function contains fast oscillations compared to all other
  relevant energy scales. Upon averaging over those fast oscillations the
  intermediate currents loose their correlations and thus the effect of
  nonlocal heat transport vanishes. All currents become equilibrium. For finite
  $W$ one observes modulations/resonances of the heat flux, which cannot change
  the sign of the corrections, but arise naturally since there is a modulation
  of the interference between the charge fluctuations in the respective
  reservoirs, mediated by the nonlocal interaction.}\BibitemShut {Stop}%
\bibitem [{Note3()}]{Note3}%
  \BibitemOpen
  \bibinfo {note} {The new $3\times 3$ scattering matrix connecting the
  incoming state and sources to the outgoing state and the two internal states
  of the resistors remains unitary. This fixes the temperature of the
  reservoirs to be equilibrium, since from the unitarity of the scattering
  matrix immediately follows that $J^c = J^c_{\protect \text {out}}$ and the
  same for the primed variables. This means that the Langevin source $j^c$, the
  current that is dissipated in the ohmic contact $j^c_{\protect \text {out}}$
  and the incoming current $j_{\protect \text {in}}$ all have a noise power
  with the same equilibrium temperature.}\BibitemShut {Stop}%
\bibitem [{Note4()}]{Note4}%
  \BibitemOpen
  \bibinfo {note} {Studying out of equilibrium situations like connecting the
  circuit or later the transmission line to a hot contact is an interesting and
  open question. However, we want to address the situation after equilibration
  has taken place and postpone the nonequilibrium and steady state properties
  of the circuit to a later point.}\BibitemShut {Stop}%
\bibitem [{Note5()}]{Note5}%
  \BibitemOpen
  \bibinfo {note} {Before tracing out the sources, the system can be understood
  as a Hamiltonian system, see \protect \cref {eq:Ham}, where different nodes
  interact according to a general capacitance matrix involving self- and
  cross-capacitances. This matrix is required to be positive definite by
  construction, which allows us to find the corresponding allowed value of
  $\lambda $.}\BibitemShut {Stop}%
\bibitem [{Note6()}]{Note6}%
  \BibitemOpen
  \bibinfo {note} {The transmission probability is only frequency independent
  if one considers the tunneling of free fermions. In leading order of the
  tunneling, the collective mode and source contribution are uncorrelated,
  which justfies this approach. Our conclusion holds for $ \protect \mathcal
  {T}$ close to $0$ or $1$ and needs to be rechecked for arbitrary $\protect
  \mathcal {T}$ . This will be considered elsewhere.}\BibitemShut {Stop}%
\bibitem [{\citenamefont {Roussel}\ \emph {et~al.}(2017)\citenamefont
  {Roussel}, \citenamefont {Cabart}, \citenamefont {Fève}, \citenamefont
  {Thibierge},\ and\ \citenamefont {Degiovanni}}]{roussel_electron_2017}%
  \BibitemOpen
  \bibfield  {author} {\bibinfo {author} {\bibfnamefont {B.}~\bibnamefont
  {Roussel}}, \bibinfo {author} {\bibfnamefont {C.}~\bibnamefont {Cabart}},
  \bibinfo {author} {\bibfnamefont {G.}~\bibnamefont {Fève}}, \bibinfo
  {author} {\bibfnamefont {E.}~\bibnamefont {Thibierge}},\ and\ \bibinfo
  {author} {\bibfnamefont {P.}~\bibnamefont {Degiovanni}},\ }\href
  {https://doi.org/10.1002/pssb.201600621} {\bibfield  {journal} {\bibinfo
  {journal} {physica status solidi (b)}\ }\textbf {\bibinfo {volume} {254}},\
  \bibinfo {pages} {1600621} (\bibinfo {year} {2017})}\BibitemShut {NoStop}%
\bibitem [{\citenamefont {Litinski}\ \emph {et~al.}(2017)\citenamefont
  {Litinski}, \citenamefont {Brouwer},\ and\ \citenamefont
  {Filippone}}]{litinski_interacting_2017}%
  \BibitemOpen
  \bibfield  {author} {\bibinfo {author} {\bibfnamefont {D.}~\bibnamefont
  {Litinski}}, \bibinfo {author} {\bibfnamefont {P.~W.}\ \bibnamefont
  {Brouwer}},\ and\ \bibinfo {author} {\bibfnamefont {M.}~\bibnamefont
  {Filippone}},\ }\href {https://doi.org/10.1103/PhysRevB.96.085429} {\bibfield
   {journal} {\bibinfo  {journal} {Physical Review B}\ }\textbf {\bibinfo
  {volume} {96}},\ \bibinfo {pages} {085429} (\bibinfo {year}
  {2017})}\BibitemShut {NoStop}%
\bibitem [{\citenamefont {Idrisov}\ \emph {et~al.}(2018)\citenamefont
  {Idrisov}, \citenamefont {Levkivskyi},\ and\ \citenamefont
  {Sukhorukov}}]{idrisov_dephasing_2018}%
  \BibitemOpen
  \bibfield  {author} {\bibinfo {author} {\bibfnamefont {E.~G.}\ \bibnamefont
  {Idrisov}}, \bibinfo {author} {\bibfnamefont {I.~P.}\ \bibnamefont
  {Levkivskyi}},\ and\ \bibinfo {author} {\bibfnamefont {E.~V.}\ \bibnamefont
  {Sukhorukov}},\ }\href {https://doi.org/10.1103/PhysRevLett.121.026802}
  {\bibfield  {journal} {\bibinfo  {journal} {Physical Review Letters}\
  }\textbf {\bibinfo {volume} {121}},\ \bibinfo {pages} {026802} (\bibinfo
  {year} {2018})}\BibitemShut {NoStop}%
\bibitem [{\citenamefont {Zhang}\ \emph {et~al.}(2022)\citenamefont {Zhang},
  \citenamefont {Gornyi},\ and\ \citenamefont
  {Sp{\aa}nsl\"att}}]{zhang_delta-t_2022}%
  \BibitemOpen
  \bibfield  {author} {\bibinfo {author} {\bibfnamefont {G.}~\bibnamefont
  {Zhang}}, \bibinfo {author} {\bibfnamefont {I.~V.}\ \bibnamefont {Gornyi}},\
  and\ \bibinfo {author} {\bibfnamefont {C.}~\bibnamefont {Sp{\aa}nsl\"att}},\
  }\href {https://doi.org/10.1103/PhysRevB.105.195423} {\bibfield  {journal}
  {\bibinfo  {journal} {Physical Review B}\ }\textbf {\bibinfo {volume}
  {105}},\ \bibinfo {pages} {195423} (\bibinfo {year} {2022})}\BibitemShut
  {NoStop}%
\end{thebibliography}%

%\cleardoublepage
%\onecolumngrid
%\appendix

%\input{supplementals}

\end{document}